\documentstyle[aps,preprint,epsfig,eqsecnum]{revtex}
\begin{document}
\draft
\title{A nonlinear approach to {\it NN} interactions using 
self--interacting meson fields}
\author{L. J\"ade and H.\,V. von Geramb}
\address{Theoretische Kernphysik, Universit\"at Hamburg\\
Luruper Chaussee 149, D-22761 Hamburg}
\date{April 3, 1996}
\maketitle
\begin{abstract}
Motivated by the success of models based on chiral symmetry in {\it NN}
interactions 
we investigate self-interacting scalar, pseudoscalar 
and vector meson fields and their impact for {\it NN} forces. 
We parametrize the corresponding 
nonlinear field equations and     
get analytic wavelike solutions. A probability amplitude for the 
propagation of particle states is calculated and applied in the
framework of a boson-exchange {\it NN} potential. 
Using a proper normalization of the meson fields makes all 
self-scattering amplitudes finite. 
The same normalization is able
to substitute for the phenomenological form factors used in  
conventional boson exchange potentials and thus yields an
phenomenological understanding of this part of the {\it NN} interaction. 
We find an
empirical scaling law which relates the meson self-interaction 
couplings to the pion mass and self-interaction coupling
constant. Our model yields $np$ phase shifts comparable to the 
Bonn B potential results and deuteron
properties, in excellent agreement with experimental data. 
\end{abstract}
\narrowtext
\section{Introduction}
One of the major problems of today's theoretical nuclear physics
is to
understand the relationship between the dynamics of the 
regime where baryons and mesons are observable and 
the underlying theory, QCD, which contains quark and gluon degrees 
of freedom. In this context chiral symmetry as a property of the 
QCD Lagrangian
should affect the interaction between hadrons at low and
medium energies. There are a number of QCD-inspired hadron models 
which are based on chiral symmetry \cite{ch} and which as a
common characteristic imply a nonlinear structure of the meson
Lagrangian. Despite their remarkable success 
in describing qualitative features of
hadron interactions, none of these models is able to reach the
quantitative agreement with experimental data provided by the
conventional boson-exchange (BE) or inversion potentials, which,
however, do not have any reference to chiral symmetry 
\cite{pot,inv}. Our goal now is to construct a potential model 
which lies between these extreme positions.
     
Rather than construct a chiral invariant Lagrangian 
we want to parametrize the possible 
effect of chiral symmetry using a phenomenological nonlinear term
in the meson Lagrangian which can be interpreted as a meson  
self-interaction. We solve the corresponding nonlinear
field equations analytically and apply the obtained meson
propagators in a boson exchange potential for the
nucleon-nucleon interaction. Doing so we want to test if and how
nonlinearities in the field equations have an influence on {\it NN}
interactions. This is done for each of the meson fields
which are used in the Bonn potential model to
yield a potential which is able to produce
scattering phase shifts and deuteron properties which are in good
quantitative agreement with experimental data, comparable well as
the Bonn potential \cite{Mach87}. 
It turns out that the nonlinear effects in
the meson fields can give rise to a propagator which not only
regularizes the meson self-energy diagrams but in addition can
substitute for the empirical form factors used in the Bonn
potential. Besides that we find an empirical connection between 
the phenomenological nonlinearities which ultimately may be traced 
onto some underlying symmetry.  
  
In our approach 
we use the model of solitary fields
developed by Burt \cite{Burt81}. Here the decoupled meson
field equation is parametrized by 
\begin{equation}
\label{fe}
\partial_{\mu}\partial^{\mu}\Phi+m^2\Phi+\lambda_1\Phi^{2p+1}+
\lambda_2
\Phi^{4p+1}=0,
\end{equation}
where $\Phi$ is the operator to describe the self-interacting
fields. For mesons with nonzero spin this operator is a vector in
Minkowski space. 
The parameter $p$ equals $1/2$ or $1$ to yield odd or even powered 
nonlinearities. Using this parametrization, 
Eq.\ (\ref{fe}) can be solved analytically. In principle  
there are a lot of different possible types of
self-interactions for pseudoscalar, scalar, and vector
meson fields. In this first step we will focus on the most
simple interaction which agrees with the constraint of
parity conservation and Lorentz invariance. This simplification 
makes
the model phenomenological but transparent and easy to handle. 
What we
expect and indeed is confirmed is that it is more important to 
include nonlinearities at all than to
specify their explicit form. 

This work is organized as follows.    
In the next section we introduce the model of solitary
meson fields. We will show how Eq.\ (\ref{fe}) can be solved
and how the solutions are quantized. After that, in 
Sec.\,\ref{mp}, a
probability amplitude for the propagation of solitary mesons is 
defined
and evaluated in momentum space. A proper normalization
of solitary mesons is depicted in Sec.\,\ref{pn} which guarantees
that all self-scattering
diagrams remain finite. The model then is applied in a
boson-exchange potential for {\it NN} interactions. No further 
modifications 
concerning the meson dynamics will
be made. In Sec.\,\ref{pnnp} we compare the amplitudes for meson 
propagation
used in the Bonn potential and our model using solitary
mesons. It will turn out that it is possible to substitute the  
Feynman
propagator multiplied with the form factors of the Bonn B 
potential by
the solitary meson propagator. Form factors or momentum space 
cutoff
parameters are not needed and additionally we will find an 
empirical
scaling law which relates all parameters of the meson 
self-interaction
to the pion mass and self-interaction coupling constant. 
  
After that we leave the close connection with the Bonn B
potential. The scaling law is employed in the 
one-solitary-boson-exchange potential (OSBEP)
which is introduced in Sec.\,\ref{osbep}. Only the structure 
of the potential is
adopted from the Bonn B OBEP. The meson-nucleon coupling
constants are taken as free parameters and are adjusted to fit
$np$ phase shifts and deuteron data. 
The pion self-interaction
coupling constant is also readjusted. All other meson
self-interaction coupling constants are then determined by the
scaling law. Finally, we will obtain a parameter set which
differs from the Bonn B parameters but
nevertheless yields a comparable accurate fit to $np$ phase
shifts and
deuteron properties which are in excellent agreement with
experimental data. Together with a discussion of future 
prospects the results are presented in Secs.\,\ref{results} 
and \ref{cao}. Additionally, $np$ scattering observables
have been calculated and will be published elsewhere
\cite{Jae96b}. For all observables, such as differential
cross sections, polarization and spin correlation
parameters, the results from OSBEP and Bonn B are
comparable in their reproduction of the angular
distributions. Furthermore,
calculations of $pp$ observables are underway. Here, the
inclusion of the Coulomb force requires several
modifications with respect to the $np$ potential considered
in this work. Besides the change of the nucleon and pion
mass a refit of the effective $\sigma${\it NN} coupling constant is
necessary to describe the data. When done, the results will
be published together with a $\chi^2$ for the $np$ as well
as $pp$ data from the OSBEP with respect to experimental
scattering observables. 
\section{Meson Self-Interaction}
\label{msi}
\subsection{Solitary mesons}
\label{sm}
To solve Eq.\ (\ref{fe}) one can make the ansatz 
for the interacting fields to depend on free fields
$\Phi=\Phi(\varphi)$ where $\varphi$ is a solution of the free 
Klein-Gordon equation (KGE) with meson mass $m$: 
\begin{equation}
\label{kg}
\partial_{\mu}\partial^{\mu}\varphi+m^2\varphi=0.
\end{equation}
This assumption reduces the nonlinear partial differential
equation (\ref{fe}) to a nonlinear ordinary differential equation
for $\Phi(\varphi)$,
\begin{eqnarray}
\lefteqn{\Phi^{''}(\varphi)+\frac{1}{\varphi}\Phi^{'}(\varphi)
-\frac{1}{\varphi^2}\Phi(\varphi)}\nonumber\\
 & & {}
-\frac{\lambda_1}{m^2\varphi^2}\Phi^{2p+1}(\varphi)
-\frac{\lambda_2}{m^2\varphi^2}\Phi^{4p+1}(\varphi)=0,
\label{ode}
\end{eqnarray}
which can be solved by direct integration. The solutions
are represented as a power
series in $\varphi$ \cite{Burt81}
\begin{equation}
\label{sol}
\Phi=\sum_{n=0}^{\infty}C_n^{1/{2p}}(w)\;b^n\;\varphi^{2pn+1}.
\end{equation}
These special wavelike solutions of Eq.\ (\ref{fe})
are oscillating functions which propagate with constant shape and 
velocity.  
Corresponding to the classical theory of nonlinear waves 
\cite{Whit75} 
they shall be called {\it solitary meson fields}.  
The coefficients $C^a_n(w)$ are Gegenbauer polynomials 
\cite{Arf70} defined by
\begin{equation}
\label{geg}
\frac{1}{(1-2xz+z^2)^a}=\sum_{n=0}^{\infty}C_n^a(x)z^n,
\end{equation}
$b$ and $w$ are
functions of the coupling constants and the order $p$ of the
self-interaction:
\begin{eqnarray}
b & = &
\left[\left(\displaystyle{\frac{\lambda_1}{4(p+1)m^2}}\right)^2
-\displaystyle{\frac{\lambda_2}{4(2p+1)m^2}}\right]^{1/2},
\nonumber\\
w & = & \displaystyle{\frac{1}{b}\frac{\lambda_1}{4(p+1)m^2}}.
\label{bw}
\end{eqnarray}
To quantize the solitary fields we use free 
wave solutions of Eq.\ (\ref{kg}) in a finite volume $V$
\cite{Itzz80},
\begin{equation}
\label{fs}
\varphi(x,k,\pm):=\frac{1}{\sqrt{2D_k\omega_kV}}\;a(k,\pm)\;
e^{\mp ikx},
\end{equation}
where the operators $a(k,\pm)$ annihilate quanta of positive or 
negative energy:
$$
\omega_k^2=\vec{k}\,^2+m^2.
$$ 
At this point it is important to notice that we added a factor  
$D_k$ which is an
arbitrary Lorentz invariant function of $\omega_k$. As will become 
obvious later this constant is
crucial for the proper normalization of solitary waves. 
The creation
operators for positive and negative energy quanta are related by
$$
a^{\dagger}(k,+)=a(k,-)\quad\mbox{and}\quad a^{\dagger}(k,-)=
a(k,+),
$$
which imply that it is sufficient to use creation and annihilation
operators for positive energy states only. The commutator of
these operators reads \cite{BjDr65}
\begin{equation}
\label{comm}
\big[a(k),a^{\dagger}(k^{'})\big]=\delta_{kk^{'}}.
\end{equation}
Using this relation one can construct a complete set of 
orthonormal 
eigenvectors to the number operator
$$
N_k=a^{\dagger}(k)a(k),
$$
which are
\begin{equation}
\label{pst}
|N_k\rangle = \frac{1}{\sqrt{N!}}\;{a^{\dagger}}^N(k)|0\rangle .
\end{equation}
These are the free $N$-particle states which form the basis 
to calculate matrix elements of solitary wave operators. 
The definition (\ref{fs}) determines the solitary meson  
operator (\ref{sol}) and ensures that the projection on the
one-particle state  
$$
|1\rangle\langle1|\Phi^{\dagger}|0\rangle=
\frac{1}{\sqrt{2D_k\omega_k  V}}\;a^{\dagger}(k)\;
e^{ikx}|0\rangle
$$
yields the plane wave solution for a free on-shell meson. 
The field operator for mesons with spin 1 is the direct product 
of a
Minkowski vector, which describes the spin polarization $s$,  
and a Lorentz invariant operator in the Hilbert space of the
$N$-particle states \cite{Itzz80}
\begin{equation}
\label{vmo1}
\Phi^{\mu}_v(x,k,s)=\epsilon^{\mu}(k,s)\phi(x,k,s).
\end{equation}
Because of the negative intrinsic parity of vector mesons, the operator 
$\phi(x,k,s)$ transforms like a pseudoscalar field:
$$
\phi\stackrel{\cal P}{\to}-\phi.
$$
To conserve parity the field equation for self-interacting vector
mesons must contain only odd powers of the fields, which restricts 
$p$ to
integer values. Using an expansion of the Proca equation 
\cite{Itzz80}  
yields
\begin{equation}
\label{vmfe}
\partial_{\nu}\partial^{\nu}\Phi^{\mu}_v+m^2\Phi_v^{\mu}+\lambda_1
(\Phi_{v}^{\nu})^{2p}\Phi^{\mu}_v+\lambda_2
(\Phi_{v}^{\nu})^{4p}\Phi^{\mu}_v=0.
\end{equation}
The normalization \cite{Itzz80}
$$
\epsilon_{\nu}(k,s)\epsilon^{\nu}(k,s^{'})=\delta_{ss^{'}}
$$
leads to the field equation for the operator $\phi$:
$$
\partial_{\nu}\partial^{\nu}\phi+m^2\phi+
\lambda_1\phi^{2p+1}+
\lambda_2\phi^{4p+1}=0.
$$
This is the same equation as Eq.\ (\ref{fe}) with integer $p$.
 Therefore
vector mesons can be described by the product of the polarization 
vector
and a solution (\ref{sol}) of the pseudoscalar field equation
\begin{equation}
\label{vmo}
\Phi_v^{\mu}(x,k,s)=\epsilon^{\mu}(k,s)\cdot\Phi_{ps}(x,k,s).
\end{equation}
\section{Meson Propagation}
\label{mp}
Using the solutions (\ref{sol}) and the commutator (\ref{comm})
for the operators appearing in Eq.\ (\ref{fs}) one can readily 
calculate matrix
elements of solitary field operators between $N$- and 
$M$-particle
states. The
probability for the propagation of an interacting field can now 
be defined as the amplitude to create an interacting system
at some space-time point $x$ which is annihilated into the vacuum
 at
$y$. Since the intermediate state is not observable and the 
particles
are not distinguishable, a weighted sum over all intermediate 
states has to be performed \cite{Jae95}:
\begin{eqnarray}
iP(y-x) & = & \sum_k\sum_{N=0}^{\infty}\displaystyle{\frac{1}{N!}}
\Big[\langle
0|\Phi(y,k)|N_k\rangle \langle
N_k|\Phi^{\dagger}(x,k)|0\rangle\theta(y_0-x_0)\nonumber\\
 & & {}+\langle 0|\Phi(x,k)|N_k\rangle \langle
N_k|\Phi^{\dagger}(y,k)|0\rangle\theta(x_0-y_0)\Big].
\label{msmp}
\end{eqnarray}
A straightforward calculation yields the desired amplitudes in
coordinate and momentum space. Defining
\begin{equation}
\label{defip}
iP(y-x)=\frac{i}{(2\pi)^4}\int d^4k \;P(k^2,m)\;e^{-ik(y-x)},
\end{equation}
the momentum space amplitude reads
\begin{eqnarray}
\lefteqn{iP(k^2,m)=\sum_{n=0}^{\infty}\Big[C_n^{1/{2p}}(w)\Big]^2}
\nonumber\\ & & \times\frac{b^{2n}}{(2V)^{2pn}}\frac{(2pn+1)
^{2pn-2}}
{D_k^{2pn+1}(\vec{k}\,^2+M_n^2)^{pn}}\;i\Delta_F(k^2,M_n),
\label{pprop1}
\end{eqnarray}
with the Feynman propagator
\begin{equation}
\label{fprop}
i\Delta_F(k^2,M_n)=\frac{i}{k_{\mu}k^{\mu}-M_n^2},
\end{equation}
and a mass-spectrum
$$
M_n=(2pn+1)m.
$$
Since $V\omega_k $ is a Lorentz scalar, the amplitude 
(\ref{pprop1}) is
Lorentz invariant. At this point it is convenient to introduce 
the  
dimensionless coupling constants $\alpha_1$ and $\alpha_2$ which we
define as
\begin{eqnarray}
\alpha_1 & := & \frac{\lambda_1}{4(p+1)m^2(2mV)^p}, \nonumber\\
 & & \nonumber\\
\alpha_2 & := & \frac{\lambda_2}{4(2p+1)m^2(2mV)^{2p}}.
\label{a1a2}
\end{eqnarray}
This yields
\begin{equation}
\label{wa}
w=\frac{\alpha_1}{\sqrt{\alpha_1^2-\alpha_2}}
\end{equation}
and
\begin{eqnarray}
iP(k^2,m) & = & \sum_{n=0}^{\infty}\Big[C_n^{1/{2p}}(w)\Big]^2 
\label{pprop}\\ 
& & \times\frac{\left[(m^p\alpha_1)^2-m^{2p}\alpha_2\right]^n 
(2pn+1)^{2pn-2}}{D_k^{2pn+1}(\vec{k}\,^2+M_n^2)^{pn}}\nonumber\\
& & \times i\Delta_F(k^2,M_n).\nonumber
\end{eqnarray}
The amplitude (\ref{pprop}) shall be referred to as {\it modified 
solitary
meson propagator}. For $p=1$ one gets the amplitude for the 
propagation
of pseudoscalar fields and $p=1/2$ describes scalar particles. 
The functional form, i.\,e.\ the mass spectrum
and the momentum dependence,  
is dominantly determined by the order $p$ of the
self-interaction rather than by the coupling constants
$\lambda_1$ and $\lambda_2$. To keep the model simple, 
we restrict our considerations to the case $\lambda_2=0$.  
The series (\ref{pprop}) converges rapidly,  
depending on the mass, and the subsequent terms diminish by two
($\pi$) or three ($\omega$) orders of magnitude; 
in practical calculations it is sufficient to use 
$n_{\mbox{\footnotesize max}}=4$. 
  
The amplitude for vector mesons requires an additional sum over the
possible polarizations 
\begin{eqnarray}
iP_v^{\mu\nu}(y-x) & = & \sum_{k,s,N}\frac{1}{N!}\Big[\langle
0|\Phi_v^{\mu}(y,k,s)|N_k\rangle \langle
N_k|{\Phi_v^{\nu}}^{\dagger}(x,k,s)|0\rangle\theta(y_0-x_0)
\nonumber\\
 & & {}+\langle 0|\Phi_v^{\mu}(x,k,s)|N_k\rangle \langle
N_k|{\Phi_v^{\nu}}^{\dagger}(y,k,s)|0\rangle\theta(x_0-y_0)\Big]
\nonumber.
\end{eqnarray}
Using the representation (\ref{vmo}) and the expression for the 
sum over the polarizations \cite{Itzz80},
\begin{equation}
\label{mt}
\sum_s\epsilon^{\mu}(k,s)\epsilon^{\nu}(k,s)=-g^{\mu\nu}+
\frac{k^{\mu}k^{\nu}}
{m_v^2},
\end{equation}
the momentum space propagator is found to be 
\begin{eqnarray}
\lefteqn{iP_{v}^{\mu\nu}(k^2,m_v)=\sum_n\left(-g^{\mu\nu}+
\frac{k^{\mu}k^{\nu}}{M_{n,v}^2}\right)(m_v\alpha_{v})^{2n}} 
\nonumber\\
 & & \times\frac{(2n+1)^{2n-2}} 
{D_{k,v}^{2n+1}\left(\vec{k}\,^2+M_{n,v}^2\right)^n}\;
i\Delta_F(k^2,M_{n,v}),
\label{vprop1}
\end{eqnarray}
where $iP_{ps}(k^2,m_v)$ equals the propagator (\ref{pprop}) with 
integer $p$ and $M_{n,v}=(2n+1)m_v$. For application in NN scattering,
one drops the part proportional to $k^{\mu}k^{\nu}$ and the propagator
simplifies to 
$$
iP_{v}^{\mu\nu}(k^2,m_v)=-g^{\mu\nu}iP_{ps}(k^2,m_v).
$$
\section{Proper Normalization}
\label{pn}
The propagator (\ref{pprop}) contains the arbitrary constant 
$D_k$. This constant can depend on the energy $\omega_k$ and the
coupling constants and is fixed by the following conditions
\cite{Burt81}: (i) all amplitudes must be Lorentz invariant, 
(ii) $D_k$ must be dimensionless, 
(iii) all self-scattering diagrams must be finite, 
(iv) the fields have to vanish for $\lambda_1,
\lambda_2\to 0$.  
Whereas the first three conditions are evident, the last 
one requires us to recall that, 
if a particle has no interaction then there is no way to
create or measure it and the amplitude for such a process
vanishes; the field exists solely because of its 
interaction.
  
The proper normalization constant is a powerful
tool to avoid the problem of regularization which arises in 
conventional 
models. In a $\lambda\Phi^4$ theory, for example, which is
described by setting $\lambda_2=0$ and $p=1$, one gets infinite 
results
calculating the first correction to the two-point function
$iP(y-x)$. A proper normalization, i.\,e., using the smallest power
$$
D_k\sim(\omega_k  V)^2,
$$
makes the result finite. A different situation occurs in models 
including massive spin-1 bosons. Such a case, with or without
self-interaction, is harder to regularize due to the additional 
momentum
dependence in the Minkowski tensor which appears in 
Eq.\ (\ref{vprop1}). 
Nevertheless, the vector mesons
$\rho$ and $\omega$ are important ingredients in every 
boson-exchange
model. A minimum power proper normalization to solve this
problem is 
$$
D_k\sim(\omega_k  V)^4.
$$
In summary, we satisfy the above stated four conditions with
\begin{equation}
\label{dkl}
D_k=\left\{1+
\left[\left(\displaystyle{\frac{m^2}{\lambda_1}}\right)^{2/p}+
\left(\displaystyle{\frac{m^2}{\lambda_2}}\right)^{1/p}\right]
(\omega_kV)^2
\right\}^{\kappa}
\end{equation}
$$
\mbox{where}\quad\left\{\begin{array}{ll} \kappa
=1 &\quad\mbox{for scalar and}\\
 & \quad\mbox{pseudoscalar mesons,} \\
\kappa=2&\quad\mbox{for vector mesons.}
\end{array}\right.
$$
For later application in {\it NN} potentials we will 
consider the interaction where $\lambda_2=0$ and $p=1/2$ or
$p=1$ for scalar or pseudoscalar and vector mesons, respectively. 
Again, 
we will use the dimensionless coupling constants (\ref{a1a2})
($\alpha=\alpha_1$) to get the following 
amplitudes 
\begin{itemize}
\item[(i)] Scalar mesons:
\begin{eqnarray}
\lefteqn{iP_s(k^2,m_s)=
\sum_{n=0}^{\infty}(\sqrt{m_s}\alpha_s)^{2n}} 
\nonumber\\
 & & \times\frac{(n+1)^{n}}
{D_{k,s}^{n+1}\left(\vec{k}\,^2+M_{n,s}^2\right)^{\frac{n}{2}}}\;
i\Delta_F(k^2,M_{n,s}),
\label{sprop}
\end{eqnarray}
with $M_{n,s}=(n+1)m_s$ and
\begin{eqnarray}
\lefteqn{D_{k,s} = 1+\left(\displaystyle{\frac{m^2}
{\lambda_s}}\right)^4
(\omega_kV)^2} \nonumber\\
 & & = 
1+\frac{1}{5184(\sqrt{m_s}\alpha_s)^4}
\left(\displaystyle{\frac{\vec{k}\,^2}{(n+1)^2}}+m_s^2\right).
\label{dksc}
\end{eqnarray}
\item[(ii)] Pseudoscalar mesons:
\begin{eqnarray}
\lefteqn{iP_{ps}(k^2,m_{ps})=\sum_n(m_{ps}\alpha_{ps})^{2n}} 
\nonumber\\
 & & \times\frac{(2n+1)^{2n-2}} 
{D_{k,ps}^{2n+1}\left(\vec{k}\,^2+M_{n,ps}^2\right)^n}\;
i\Delta_F(k^2,M_{n,ps}),
\label{psprop}
\end{eqnarray}
with $M_{n,ps}=(2n+1)m_{ps}$ and
\begin{eqnarray}
\lefteqn{D_{k,ps} = 1+\left(\displaystyle{\frac{m^2}
{\lambda_{ps}}}\right)^2
(\omega_kV)^2} \nonumber\\
 & & =1+\frac{1}{256(m_{ps}\alpha_{ps})^2}
\left(\displaystyle{\frac{\vec{k}\,^2}{(2n+1)^2}}+m_{ps}^2\right).
\label{dkps}
\end{eqnarray}
\item[(iii)] Vector mesons:
\begin{eqnarray}
\lefteqn{iP_{v}^{\mu\nu}(k^2,m_v)=\sum_n\left(-g^{\mu\nu}+
\frac{k^{\mu}k^{\nu}}{M_{n,v}^2}\right)(m_v\alpha_{v})^{2n}} 
\nonumber\\
 & & \times\frac{(2n+1)^{2n-2}} 
{D_{k,v}^{2n+1}\left(\vec{k}\,^2+M_{n,v}^2\right)^n}\;
i\Delta_F(k^2,M_{n,v}),
\label{vprop}
\end{eqnarray}
with $M_{n,v}=(2n+1)m_{v}$ and
\begin{eqnarray}
\lefteqn{D_{k,v} = \left[1+\left(\displaystyle{\frac{m^2}{\lambda_v}}
\right)^2
(\omega_kV)^2\right]^2} \nonumber\\
 & & =\left[1+\displaystyle{\frac{1}{256(m_{v}\alpha_{v})^2}
\left(\displaystyle{\frac{\vec{k}\,^2}{(2n+1)^2}}+
m_{v}^2\right)}\right]^2.
\label{dkv}
\end{eqnarray}
\end{itemize}
\section{Propagator and {\it NN} Potentials}
\label{pnnp}
With its proper normalization the modified solitary meson
propagator is now completely determined. Without any modification
we next apply the
model of solitary mesons in a boson-exchange potential for {\it NN}
interactions. As guiding line we use the Bonn potential  
\cite{Mach87}. In this model different sets of parameters 
allow a good quantitative
description of {\it NN} scattering data as well as deuteron and nuclear
matter properties. Nevertheless, these potentials contain 
phenomenological
cutoff masses entering the form factors applied at each 
meson-nucleon
vertex. The cutoff
masses are free parameters and each of them is adjusted to fit
the experimental data. 
  
As a special example we regard the Bonn B OBEP  \cite{Mach89} 
since this
parametrization of the Bonn potential is easy to handle and 
provides
good agreement with experiment. The Bonn B potential takes
into account the 
contributions of mesons with masses below 1\,GeV. These are the
scalar $\delta$, the pseudoscalar $\pi$ and $\eta$, and the vector 
mesons
$\rho$ and $\omega$. Additionally one includes the virtual 
scalar $\sigma_0$ and $\sigma_1$ in isospin $T=0$ and $T=1$ 
channels, respectively, to simulate correlated two-pion
exchange \cite{Mach87}. 
  
In our approach we will adopt all features 
concerning nucleon kinematics and the structure of meson-nucleon
coupling from the Bonn B potential. 
The important difference will be the treatment of the meson
dynamics. In the Bonn B model a renormalized Feynman propagator is
used for the mesons in intermediate states. Additionally 
one has to attach a form factor to each meson-nucleon vertex to 
achieve
convergence of the momentum space potentials. Combined one gets the
amplitude for the exchange of meson $\beta=\pi$,...,$\delta$,
\begin{equation}
\label{ff}
\Big[F_{\beta}(k^2)\Big]^2i\Delta_F(k^2,m_{\beta})
\end{equation}
with
$$
F_{\beta}(k^2)=\left(\frac{\Lambda_{\beta}^2-m_{\beta}^2}
{\Lambda_{\beta}^2
-k^2}\right)^{n_{\beta}}. 
$$
For vector mesons this amplitude is multiplied with the tensor
(\ref{mt}). The values of the Bonn B parameter set for the 
cutoff masses 
and the exponent $n_{\beta}$ can be found in Table \ref{bbp}. 
  
In a model with solitary bosons as excess particles the modified
solitary meson propagator (\ref{pprop}) should be used instead of 
the
free Feynman propagator. The question arises if form factors are
still necessary. In this context one must remember that,  
due to the proper normalization, the solitary
meson  amplitude already contains a strong decay 
with increasing
momentum. It is important to emphasize that this normalization 
was done
independently of the application in {\it NN} interactions. Therefore if 
the amplitude (\ref{ff}) is replaced by the modified solitary meson
propagator with proper normalization we may be able to solve  
two problems of a
boson-exchange potential simultaneously: (i) 
regularization of  
self-energy amplitudes and (ii) convergence of momentum 
space  
potentials. Form factors are not necessary any more and the 
propagator  
(\ref{pprop}) is all that is needed to describe the meson 
dynamics. 
  
In the Bonn potential, form factors are used 
in a sophisticated but empirical manner. Replacing this
phenomenology by effects of nonlinearities in the meson
field equations we conjecture new aspects in this part of the
boson-exchange model.   

To test the validity of this substitution we first compare 
the
Feynman propagator multiplied with the form factor of the Bonn 
potential
to the solitary meson propagators (\ref{sprop}-\ref{vprop}) 
before we
calculate a solitary boson exchange potential. 
  
To make this comparison we recall that we use the  
Blan\-ken\-becler-Sugar (BbS) reduction \cite{BbS}
of the Bethe-Salpeter (BS) equation \cite{BS}. Here the 
relativistic  
two-nucleon propagator is replaced by the BbS propagator
$$
g(k,s)=\delta(k_0)\;\frac{M^2}{E_k}\;
\frac{\Lambda^{(1)}_+
(\vec{k})\Lambda^{(2)}_+(-\vec{k})}{\frac{1}{4}s-E_k^2+i\epsilon},
$$
where $\sqrt{s}$ is the total energy in the center-of-mass frame, 
$k$ is
the momentum of the intermediate meson, $M$ the nucleon mass and
$\Lambda^{(i)}_+$ is the positive energy projection operator for 
nucleon $i$. As a result of the delta-function, 
the matrix element of {\it NN}
scattering is applied at $k_0=0$. The amplitudes describing meson
propagation therefore depend on $\vec{k}\,^2$ only and can easily be
compared. 
   
\begin{figure} \centering
\begin{picture}(7,6.36)(0.0,0.0)
\epsfig{figure=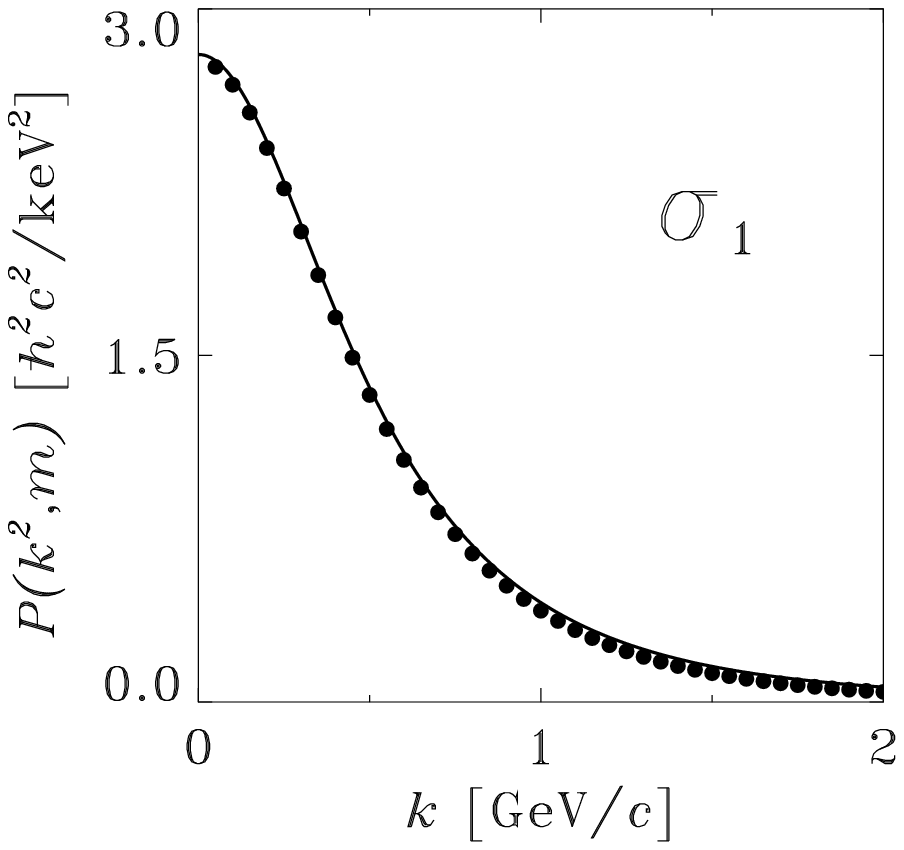,width=7.0cm}
\end{picture}
\\
\begin{picture}(7.0,6.36)(0.0,0.0)
\epsfig{figure=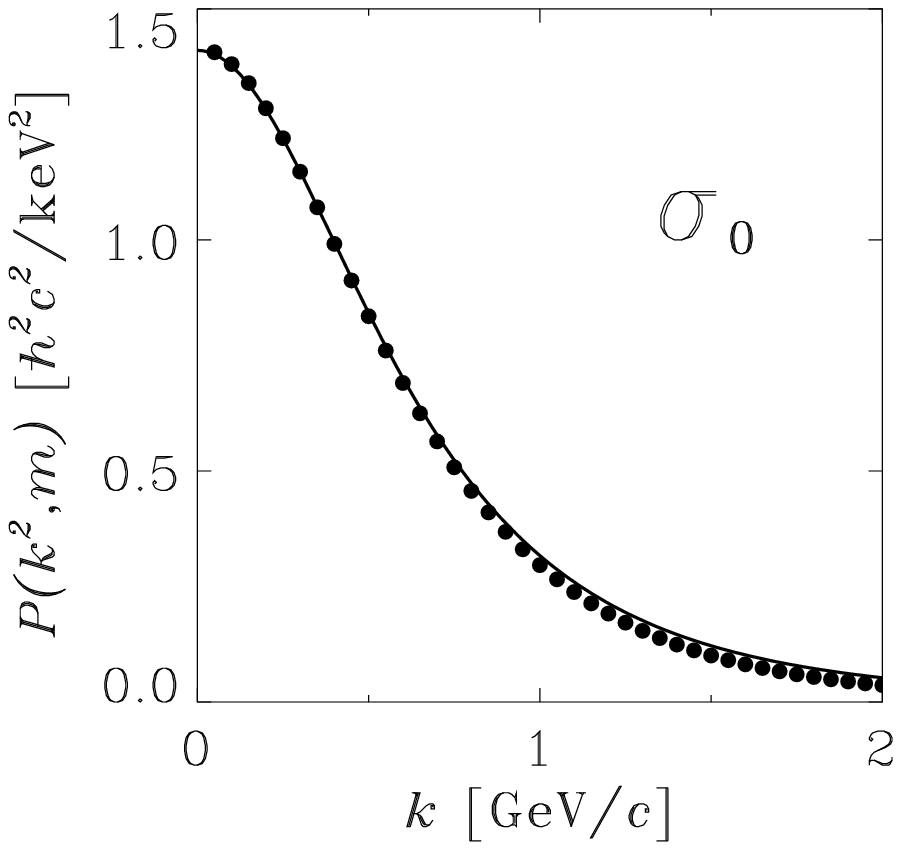,width=7.0cm}
\end{picture}
\\
\begin{picture}(7.0,6.36)(0.0,0.0)
\epsfig{figure=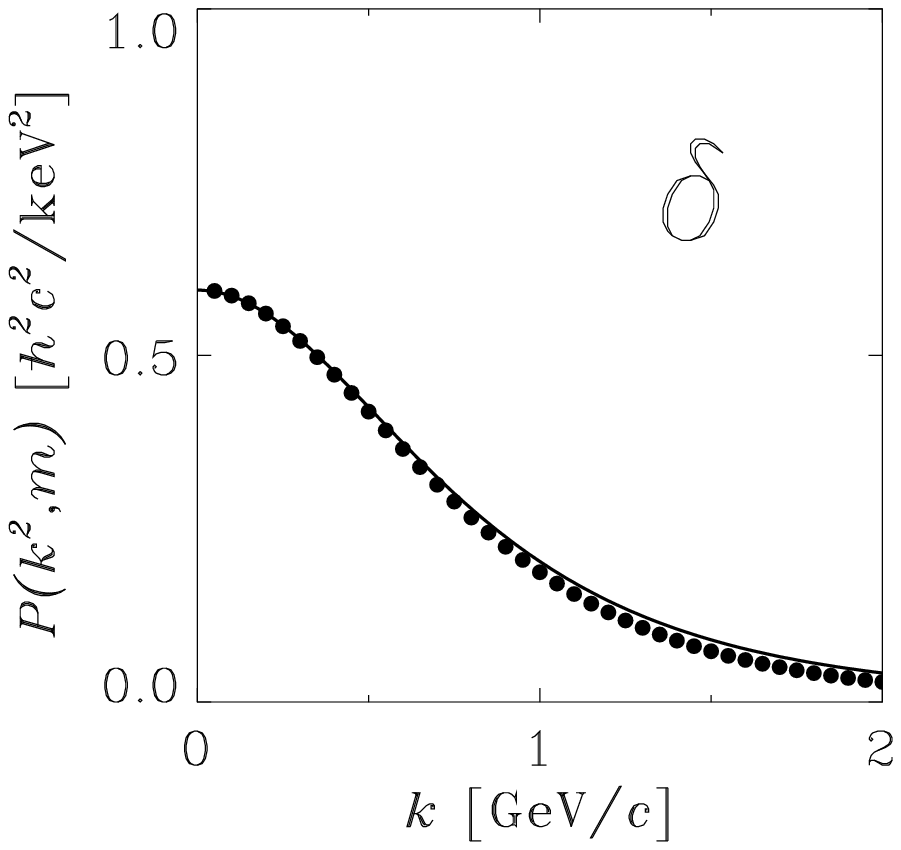,width=7.0cm}
\end{picture}
\caption[Scalar Bonn B and $\lambda\Phi^2$ propagator]
{{Solitary meson propagators for quadratic self-interaction (full) 
compared to the corresponding Feynman-propagators multiplied with the
Bonn B form factors (circles). \label{pnscprop}}}
\end{figure}

\begin{figure} \centering
\begin{picture}(7,6.36)(0.0,0.0)
\epsfig{figure=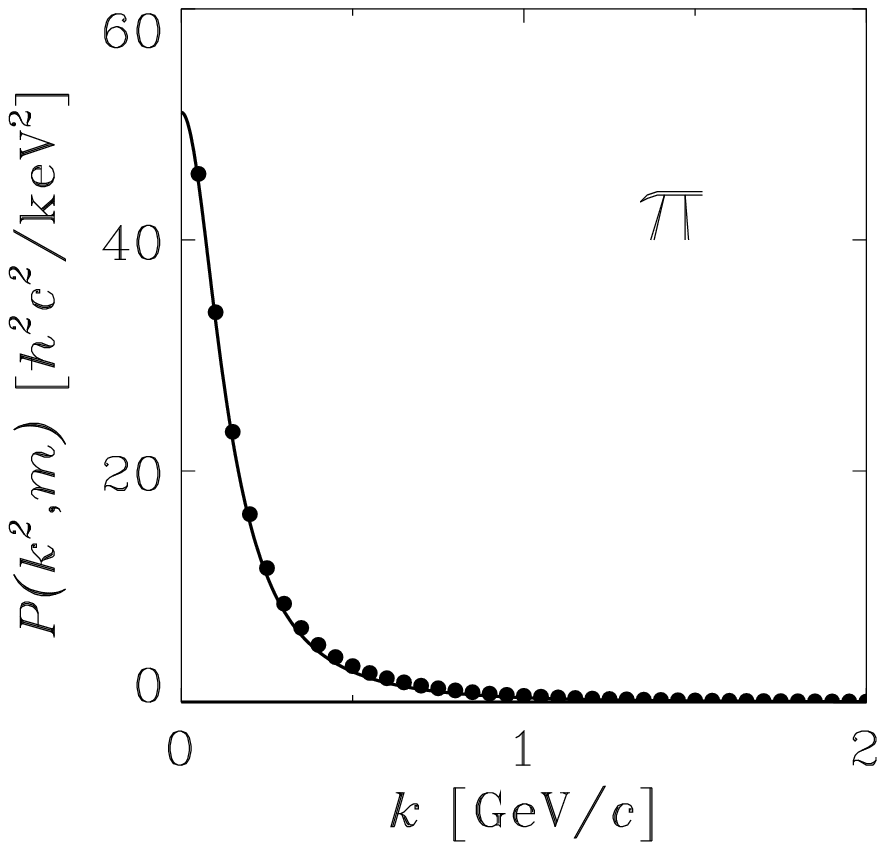,width=7cm}
\end{picture}
\begin{picture}(7,6.36)(0.0,0.0)
\epsfig{figure=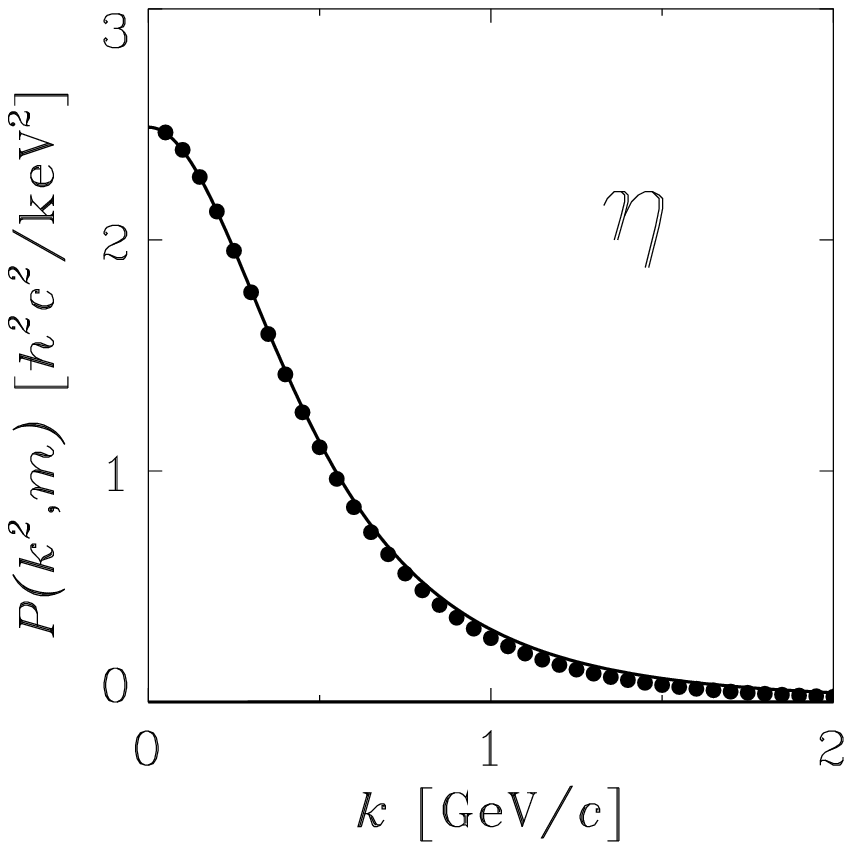,width=7cm}
\end{picture}
\\
\begin{picture}(7,6.36)(0.0,0.0)
\epsfig{figure=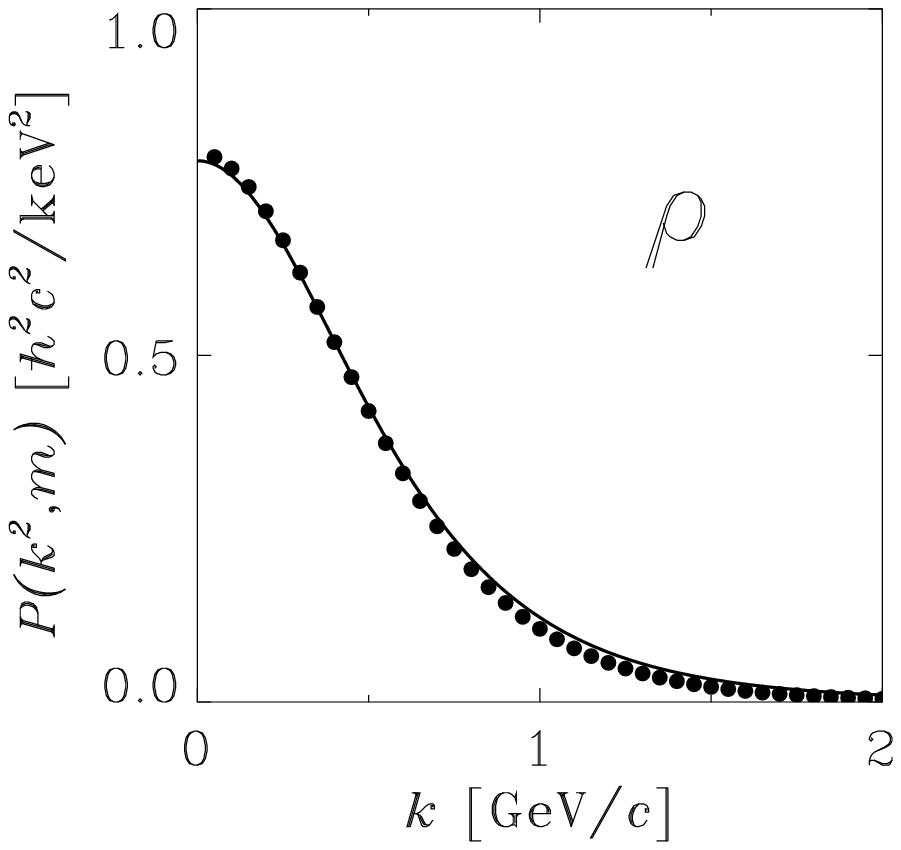,width=7cm}
\end{picture}
\begin{picture}(7,6.36)(0.0,0.0)
\epsfig{figure=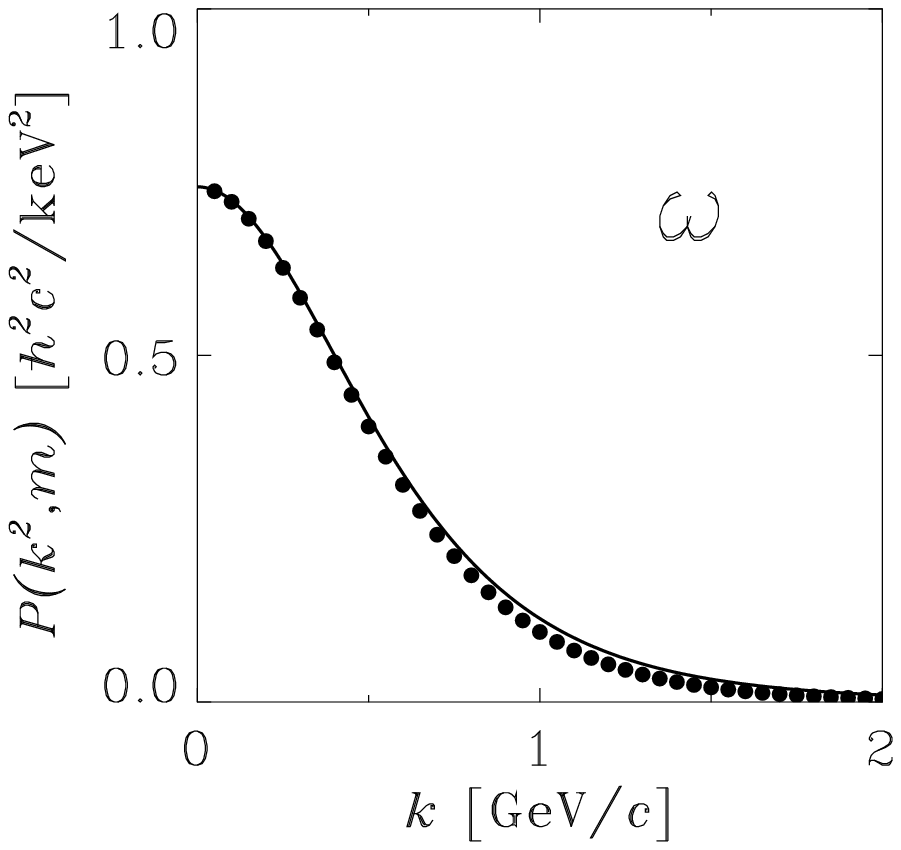,width=7cm}
\end{picture}
\caption[Bonn B and $\lambda\Phi^3$ propagators]
{{Solitary meson propagators for cubic self-int.\ (full) versus Bonn B
amplitudes (circles). 
\label{pnpsprop}}}
\end{figure}

In Figs.\,\ref{pnscprop} and \ref{pnpsprop} we show the Bonn B
amplitudes (\ref{ff})
for the different mesons compared with the solitary meson
propagators. Using the parameters of the Bonn B potential
one can always find appropriate values of the self-interaction
coupling constant $\alpha_{\beta}$ to
make the solitary wave propagator fit the Feynman propagator 
multiplied by the form factor. Obviously the
solitary meson propagator yields the same amplitude as the Feynman
propagator dressed with the meson-nucleon form factor. This
confirms our anticipation that inclusion of a more elaborated meson
dynamics might be able to substitute for the form factor.  
    
\begin{figure} \centering
\begin{picture}(8.727,6)(0.0,0.0)
\epsfig{figure=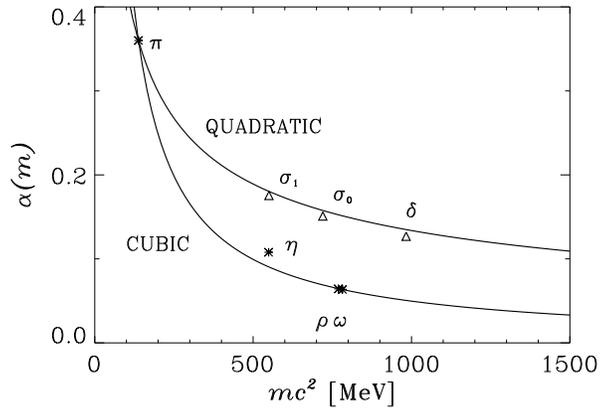,width=8.727cm}
\end{picture}
\caption[Scaling Law]
{{Coupling constants $\alpha_{\beta}$ predicted by the scaling law
(\protect\ref{scaling}) for quadratic and cubic 
self-interaction compared to the values which were found by fitting 
the Bonn B amplitudes. \label{scale}}}
\end{figure}

The coupling constants which were found by fitting the solitary 
meson
propagators to the amplitudes (\ref{ff}) are listed in Table 
\ref{sicc}.
It is obvious that they decrease with increasing
mass which implies a dependence $\alpha(m)$. In fact, looking at 
the
propagators (\ref{sprop})-(\ref{vprop}) one finds that the mass 
and the
coupling constant always appear in the combination 
$\sqrt{m}\alpha$ for
scalar and $m\alpha$ for pseudoscalar and vector fields. From 
this one
can guess an empirical scaling law which reads
\begin{equation}
\label{scaling}
\begin{array}{rrcll}
 & \alpha(m) & = &
\alpha_{\pi}\cdot
\left(\displaystyle{\frac{m_{\pi}}{m}}\right)^{\frac{1}{2}}
 & \quad\mbox{for scalar fields,}\\
 & & & & \\
 & \displaystyle{\frac{\alpha(m)}{\sqrt{\kappa}}} 
& = &
\displaystyle{\frac{\alpha_{\pi}}{\sqrt{\kappa_{\pi}}}}
\cdot\left(\displaystyle{\frac{m_{\pi}}{m}}\right)
 & \quad\mbox{for pseudoscalar} \\
 & & & & \quad\mbox{and vector fields.}
\end{array}
\end{equation}
To test these relations we took the pion self-interaction coupling
constant $\alpha_{\pi}=0.36$ and calculated all other coupling 
constants
using the scaling law (\ref{scaling}). In Fig.\,\ref{scale} we 
show the 
predicted mass dependence for fields with quadratic and cubic
self-interactions compared to the coupling constants from Table
\ref{sicc} obtained by
fitting the propagator (\ref{pprop}) to the Bonn B amplitudes. The
agreement for both kinds of self-interactions is amazing. 
Consequently, we conclude that the modified solitary meson 
propagator is
not only an
equivalent description for the meson exchange amplitude as the 
Feynman
propagator combined with a form factor but additionally yields a
connection between the meson parameters by a simple scaling
law. This might be the residue of some underlying symmetry and 
therefore
puts some physical significance to the parameter $\alpha_{\pi}$.
  
The deviations from the predicted mass dependence
are understandable since the cutoff masses 
of the Bonn B potential are free and uncorrelated
parameters. Thus their values can be 
readjusted in a global fit
together with the meson-nucleon coupling constants, enforcing  
the self-interaction coupling constants
to agree with the scaling law. 
  
In the next section we will calculate a boson-ex\-change potential
using self-interacting mesons as exchange particles. From the 
beginning
we will employ the scaling law (\ref{scaling}) and therefore the 
pion
self-interaction coupling constant will be the only parameter 
for the
dynamics of all mesons. 
\section{One-Solitary-Boson-Exchange Potential}
\label{osbep}
In the former section we have shown the equivalence between 
describing the
meson exchange amplitude by a Feynman propagator multiplied 
with a form
factor and the usage of the modified solitary meson propagators 
(\ref{sprop})-(\ref{vprop}). With this identification it is straight
forward to calculate a momentum space potential for elastic {\it NN}
scattering. We take the Bonn B potential and replace the amplitude
(\ref{ff}) by the propagator (\ref{pprop}). All other features 
remain
essentially unchanged. A detailed description of the potential 
can be
found in the genuine publication by Machleidt, Holinde, and Elster 
\cite{Mach87}; for further details, see Refs.\ \cite{Mach89,Mach93}. 
  
Since the conventional terminology of form factors is abandoned 
in the 
solitary boson-exchange potential, we do not have any cutoff
parameters in our model. We strictly use the 
scaling law for the meson self-interaction. 
Besides the meson-nucleon coupling constants the pion 
self-interaction coupling constant is the only adjustable
parameter; all other meson parameters are determined by 
the scaling law (\ref{scaling}). After adjusting the free 
parameters to
$np$ data we obtain a parameter set which differs
from the Bonn B parameters, implying that the solitary meson
propagators no longer fit the Bonn B amplitudes. However, 
the results in $np$ interactions are in very good 
agreement with
experimental data, as we show in the next section.  
\section{Results}
\label{results}
\begin{figure} \centering
\begin{picture}(7.85,5.4)(0.0,0.0)
\epsfig{figure=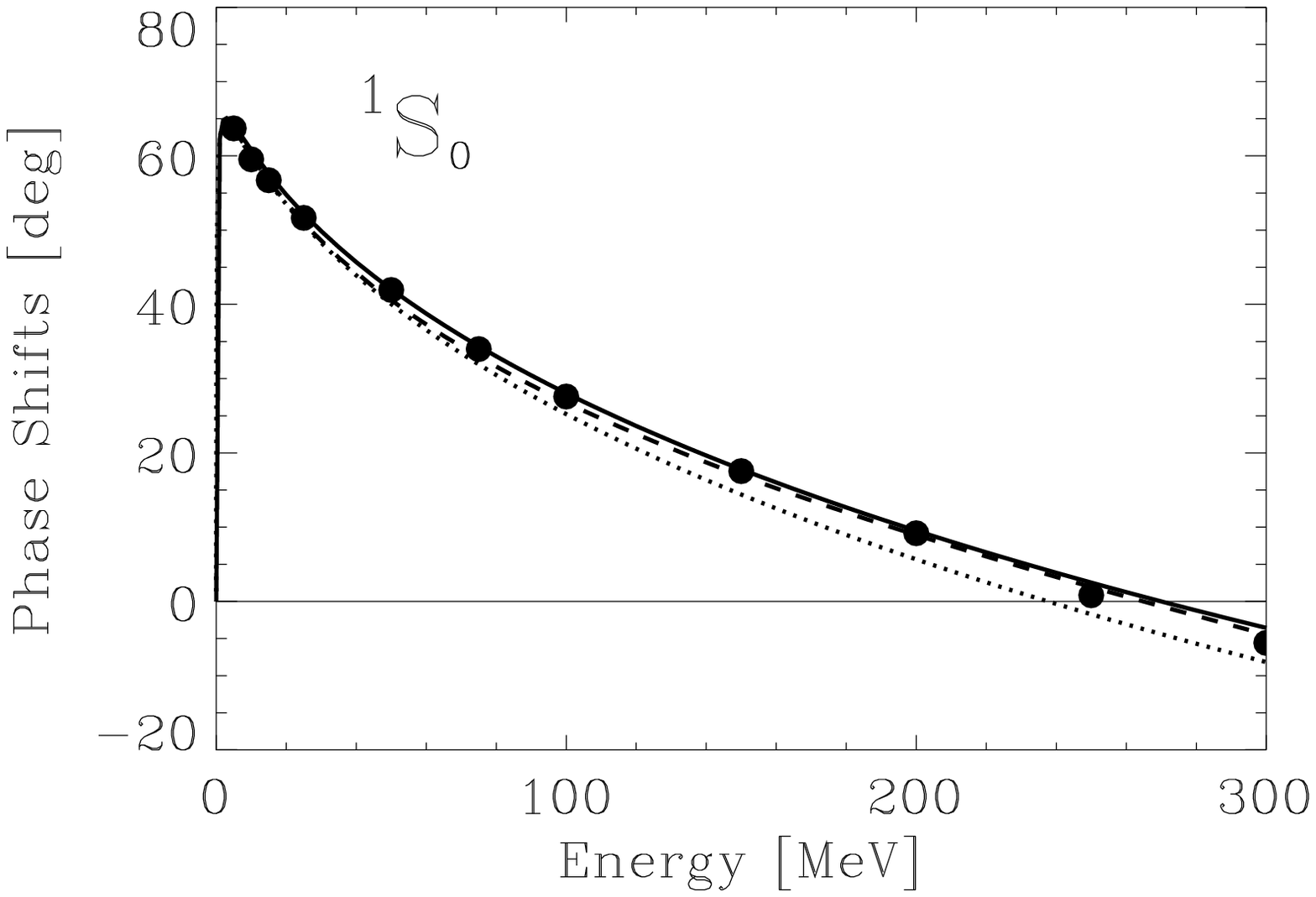,width=7.85cm}
\end{picture}
\begin{picture}(7.85,5.4)(0.0,0.0)
\epsfig{figure=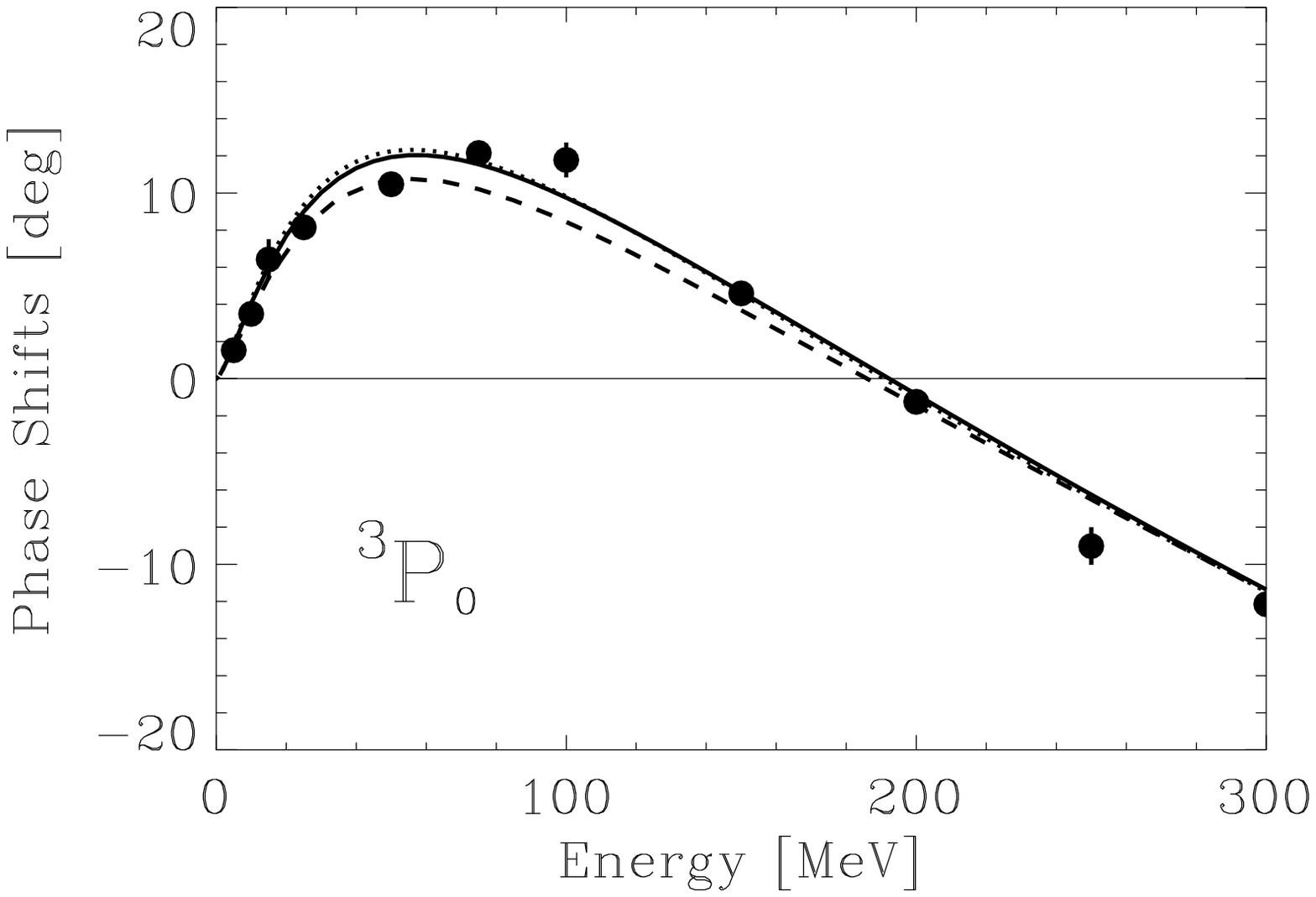,width=7.85cm}
\end{picture}
\\
\begin{picture}(7.85,5.4)(0.0,0.0)
\epsfig{figure=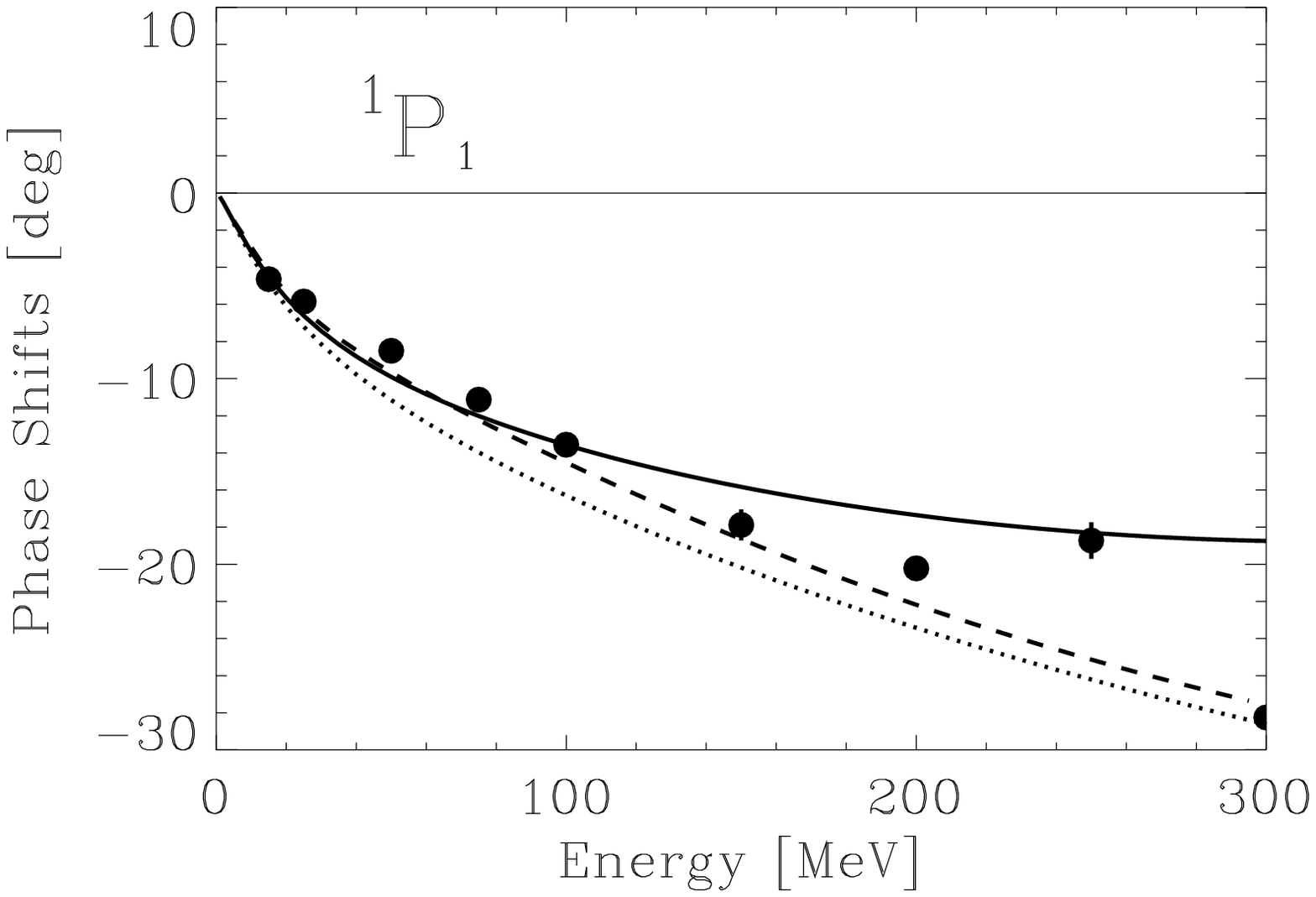,width=7.85cm}
\end{picture}
\begin{picture}(7.85,5.4)(0.0,0.0)
\epsfig{figure=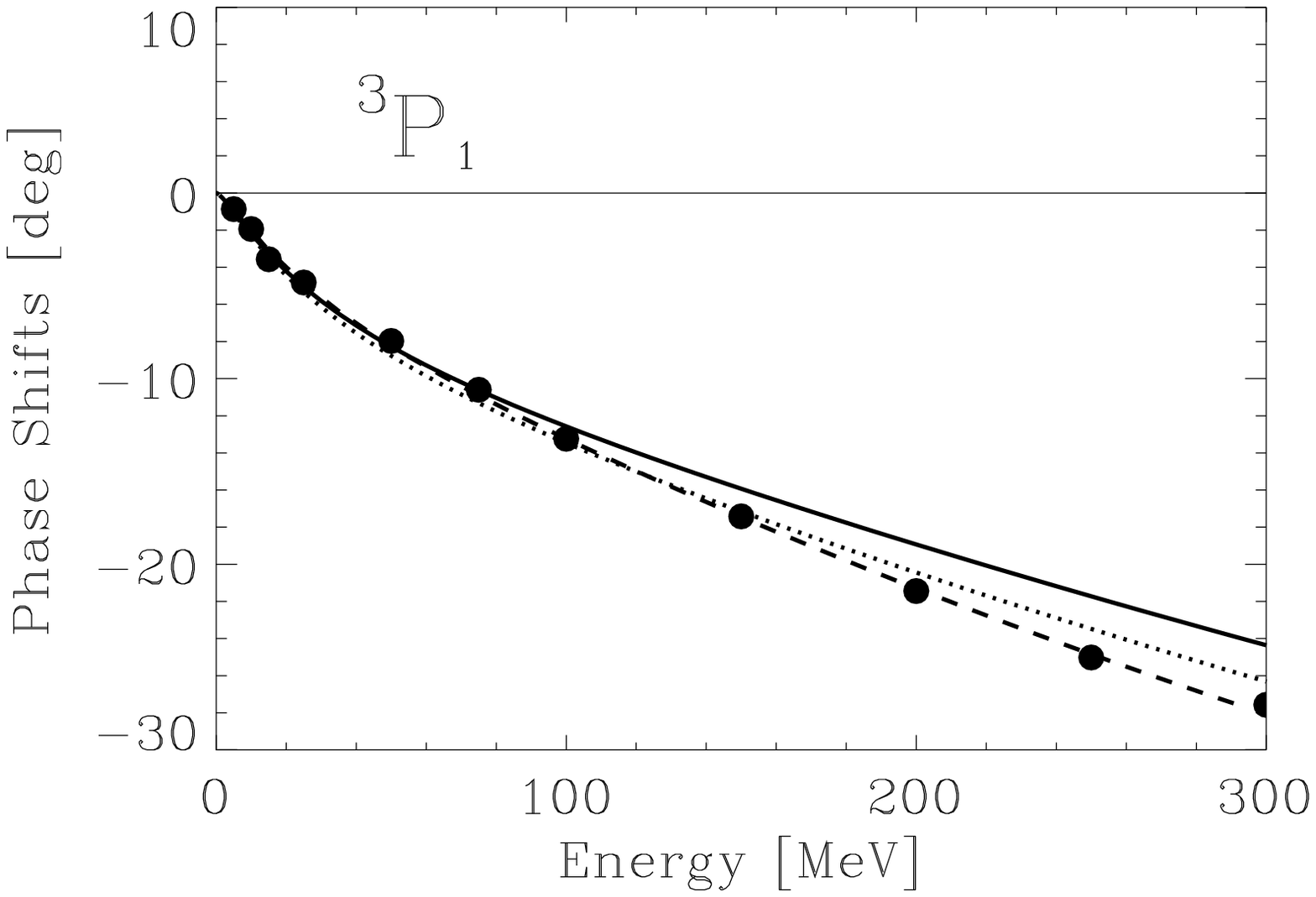,width=7.85cm}
\end{picture}
\\
\begin{picture}(7.85,5.4)(0.0,0.0)
\epsfig{figure=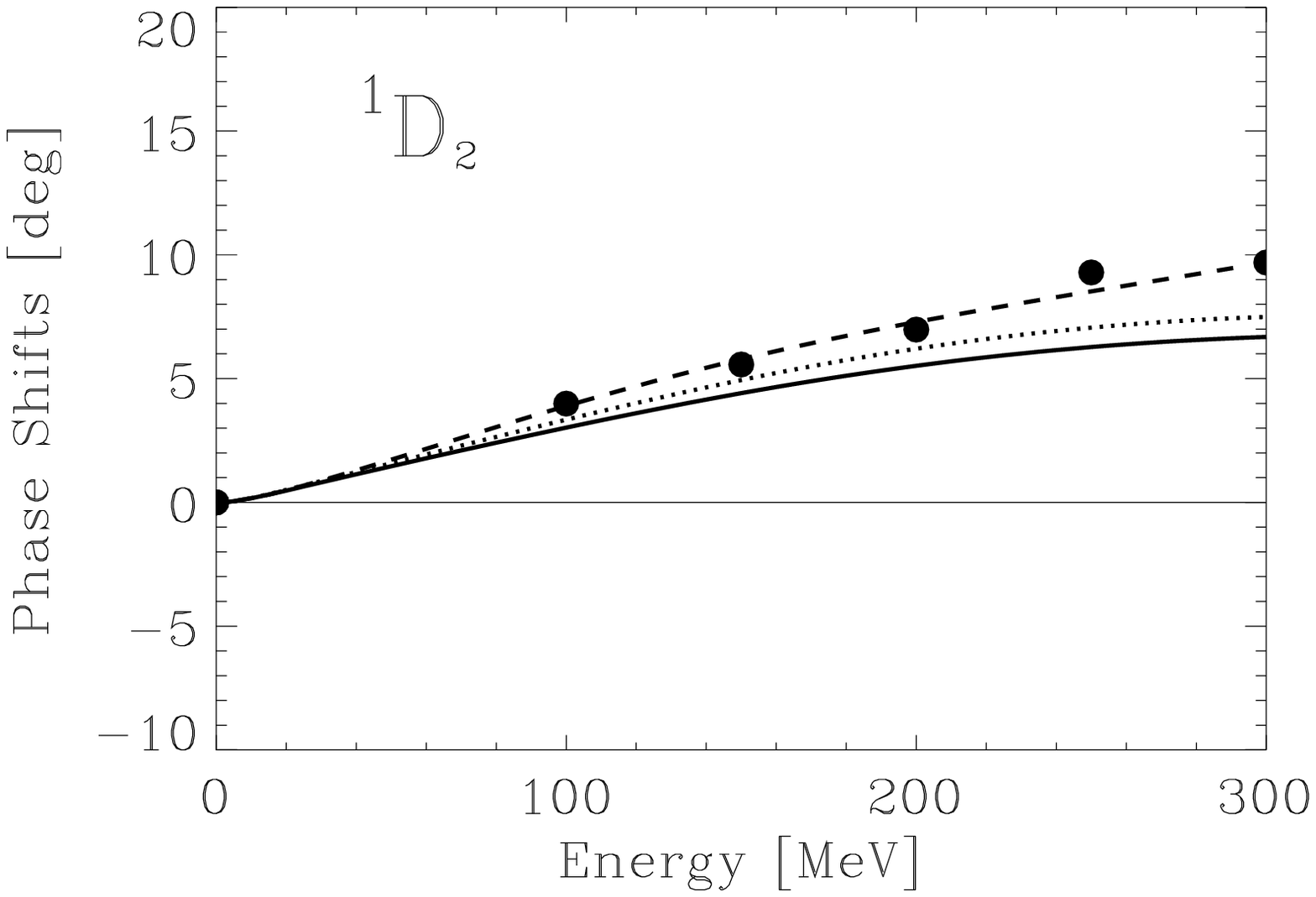,width=7.85cm}
\end{picture}
\begin{picture}(7.85,5.4)(0.0,0.0)
\epsfig{figure=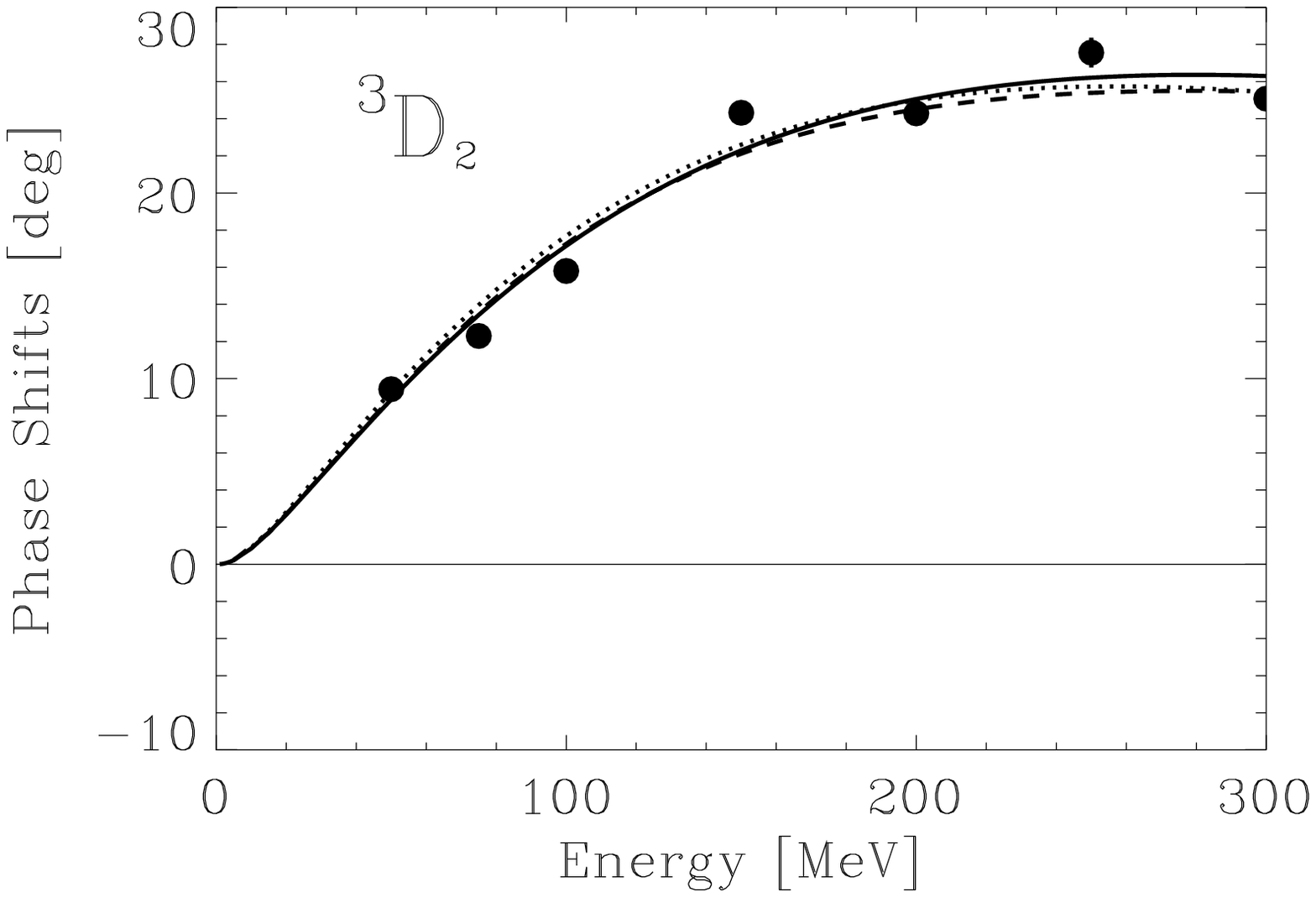,width=7.85cm}
\end{picture}
\\
\begin{picture}(7.85,5.4)(0.0,0.0)
\epsfig{figure=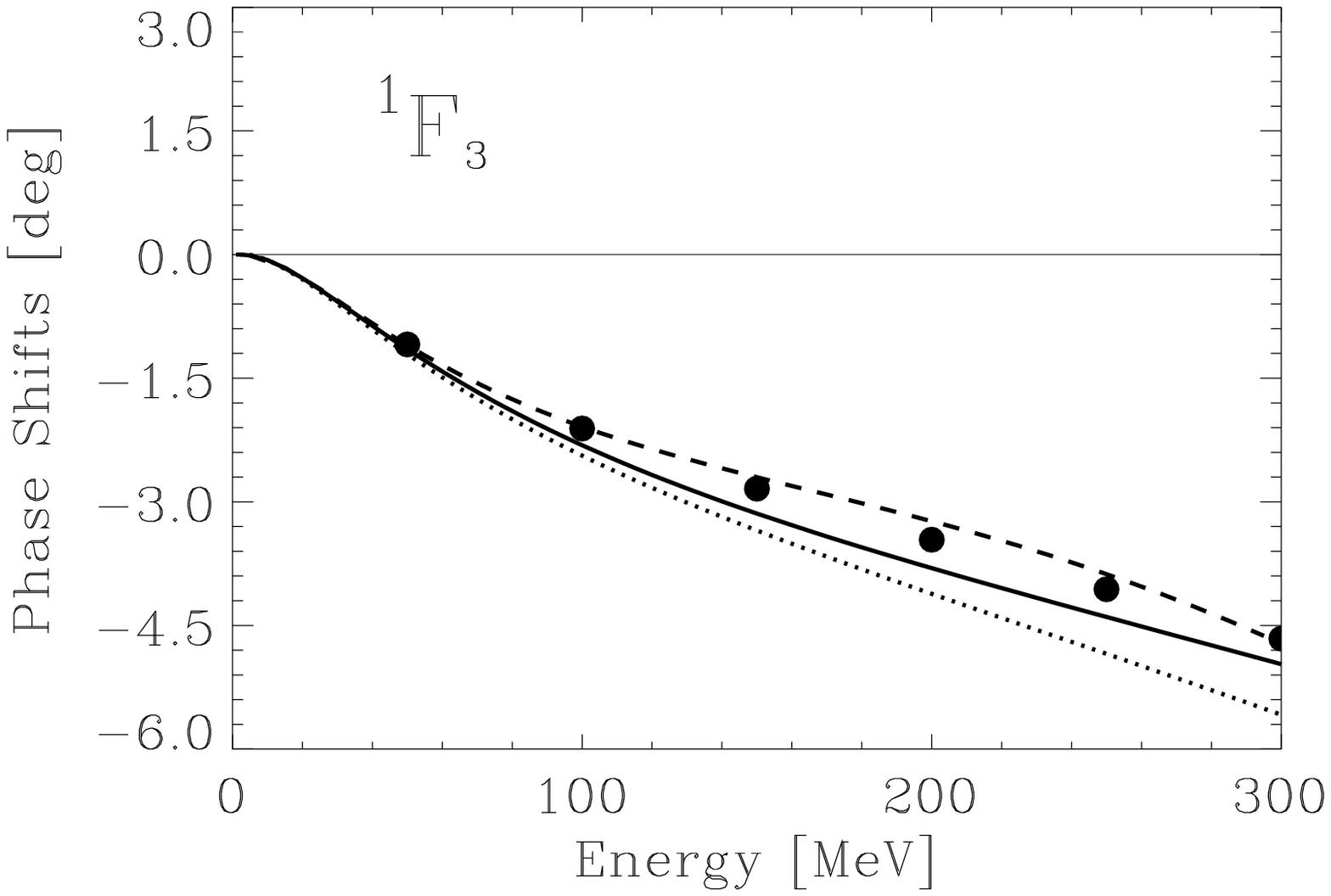,width=7.85cm}
\end{picture}
\begin{picture}(7.85,5.4)(0.0,0.0)
\epsfig{figure=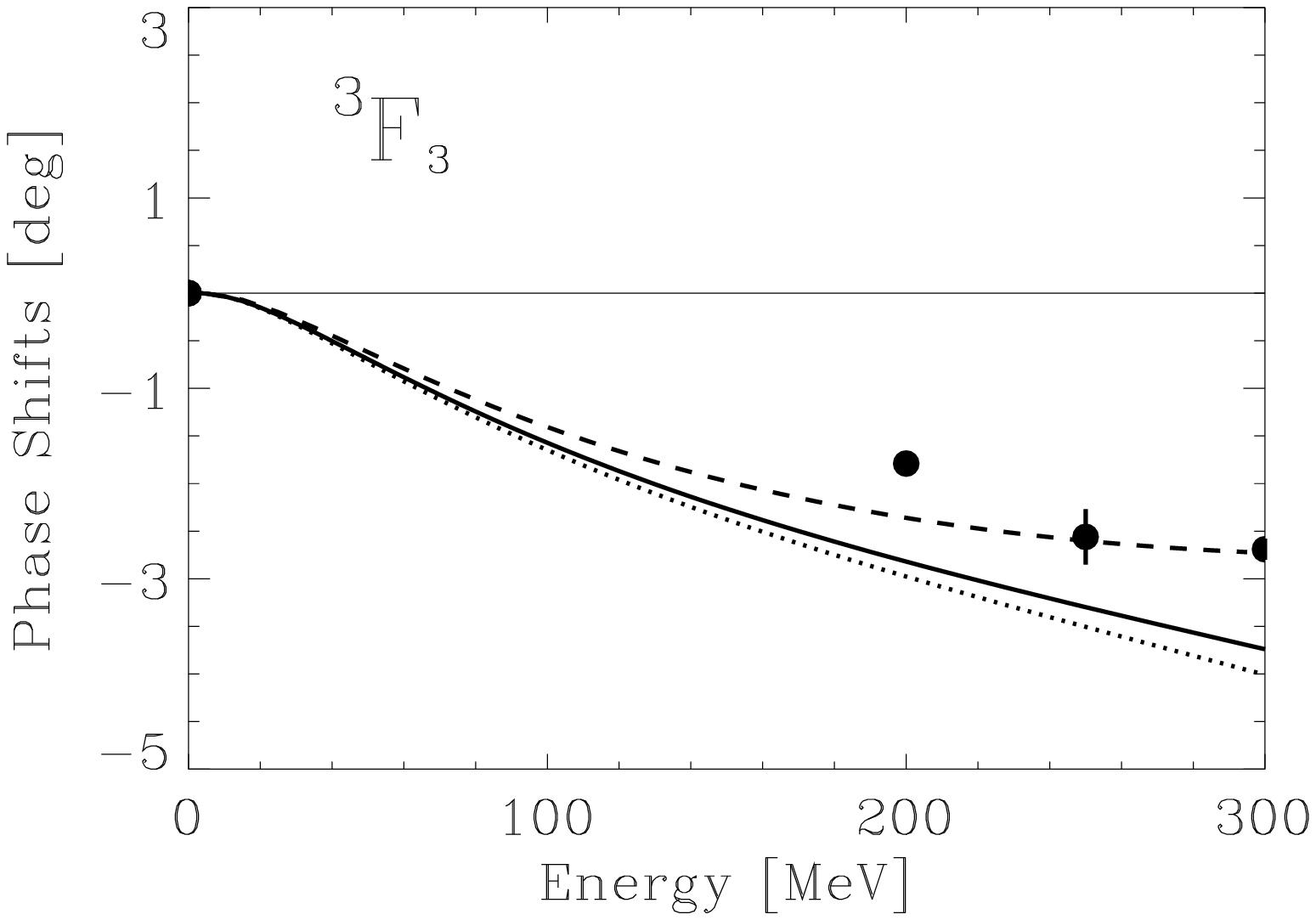,width=7.85cm}
\end{picture}
\caption[$np$ Single Channel Phase Shifts]{$np$ single channel 
phase 
shifts. We show Arndt SM95 (circles), Nijmegen PWA
(dashed), Bonn B (dotted) and OSBEP (full).}
\label{sc}
\end{figure}

\begin{figure} \centering
\begin{picture}(8,5.5)(0.0,0.0)
\epsfig{figure=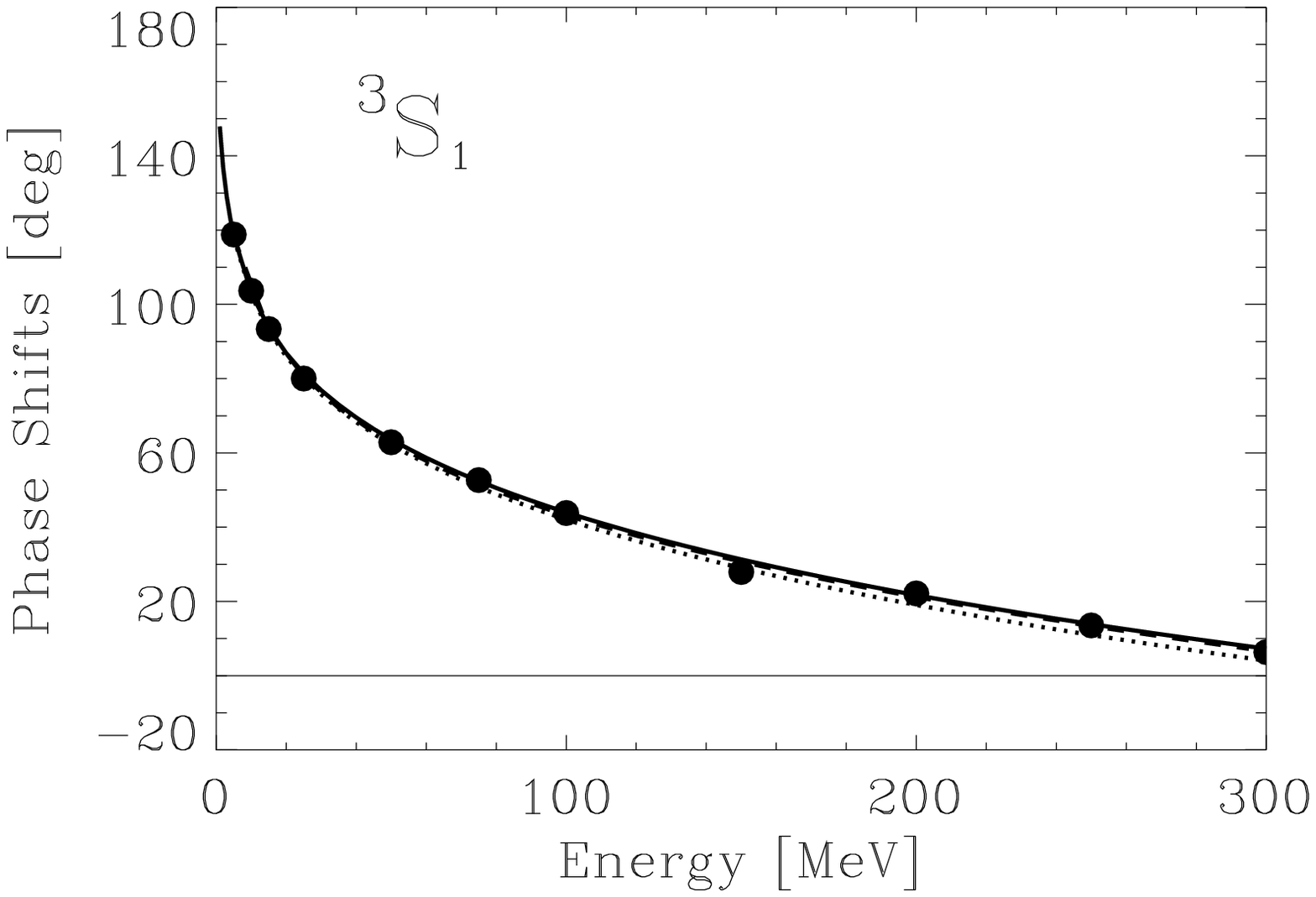,width=8.0cm}
\end{picture}
\begin{picture}(8,5.5)(0.0,0.0)
\epsfig{figure=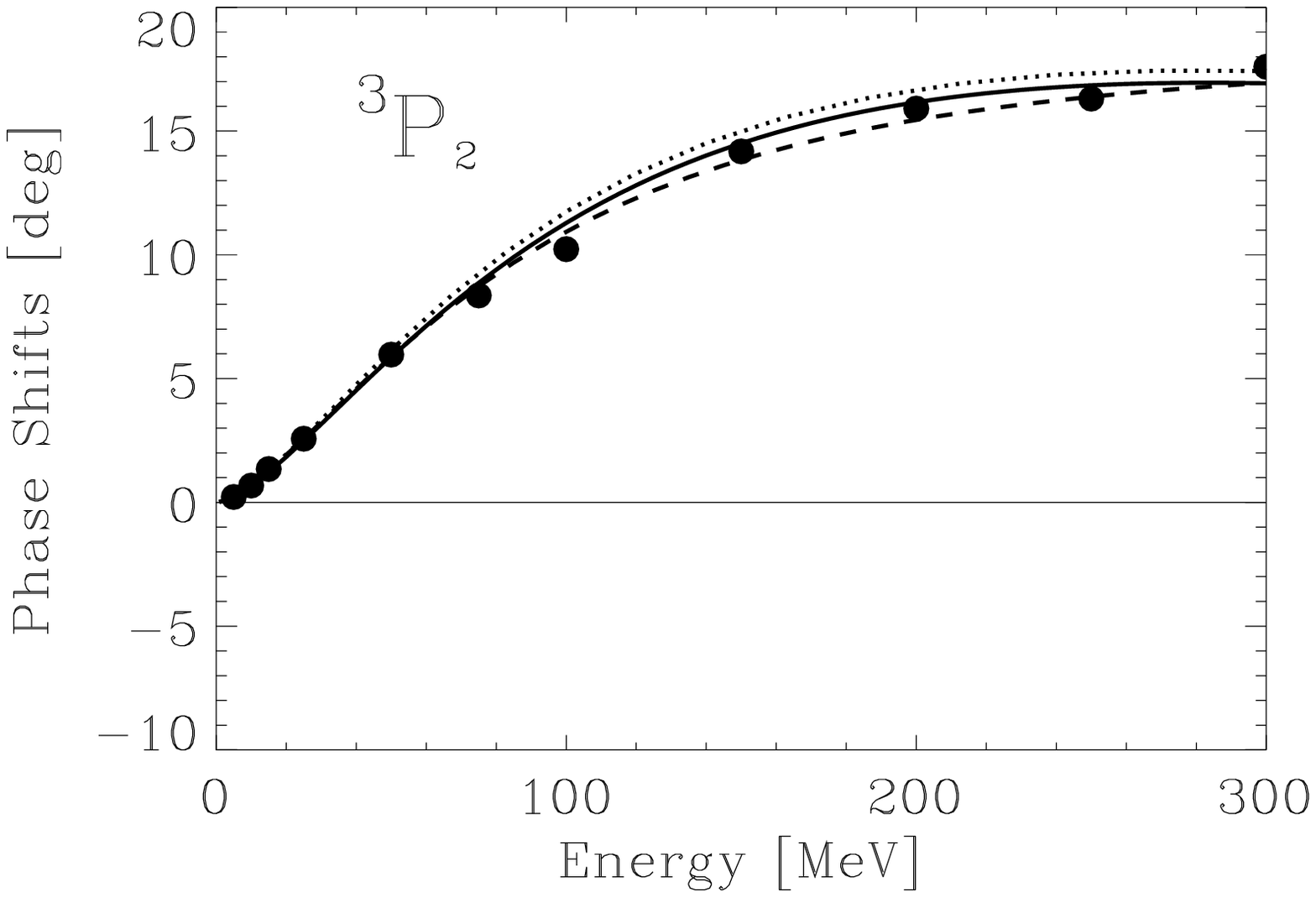,width=8.0cm}
\end{picture}
\\
\begin{picture}(8,5.5)(0.0,0.0)
\epsfig{figure=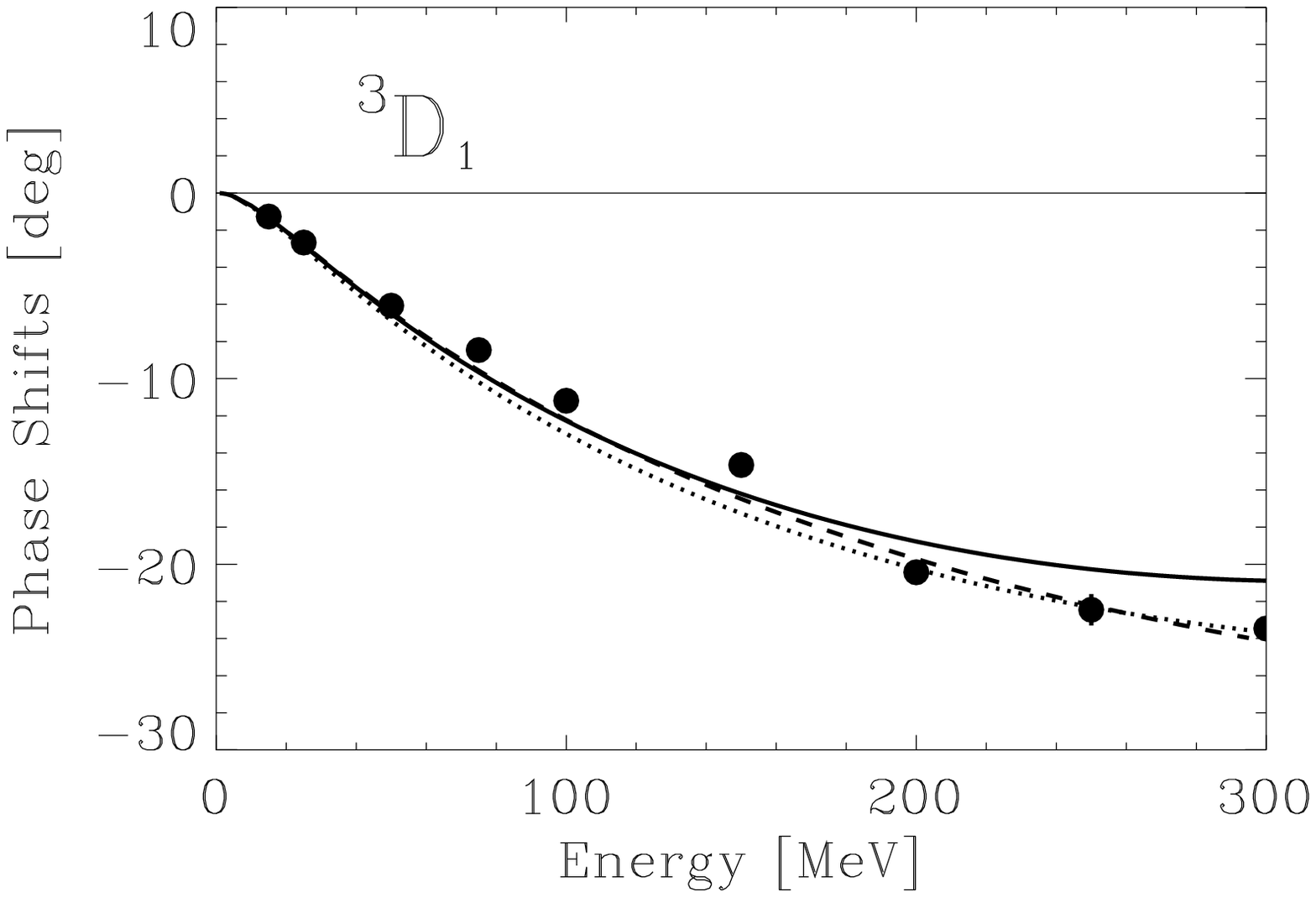,width=8.0cm}
\end{picture}
\begin{picture}(8,5.5)(0.0,0.0)
\epsfig{figure=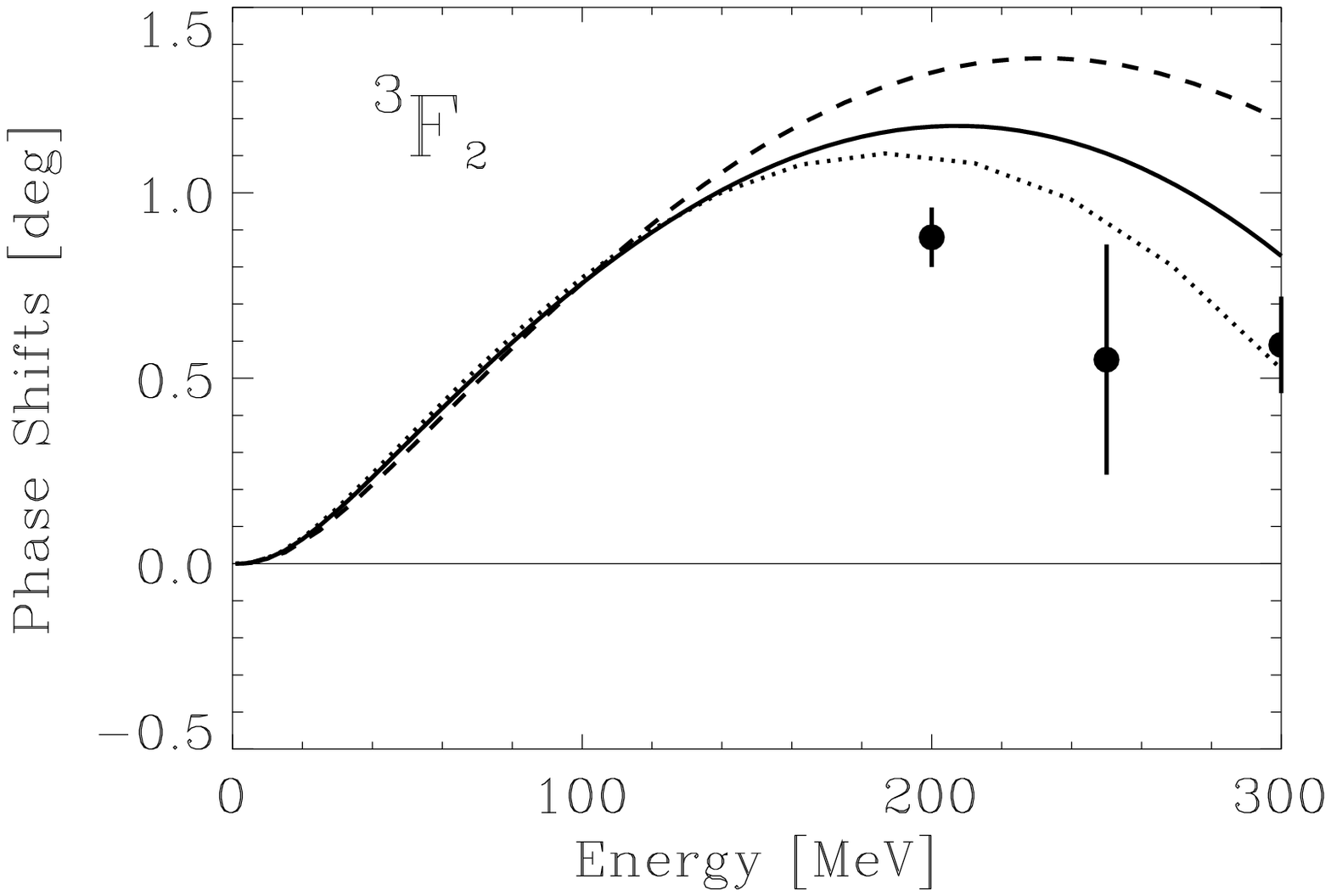,width=8.0cm}
\end{picture}
\\
\begin{picture}(8,5.5)(0.0,0.0)
\epsfig{figure=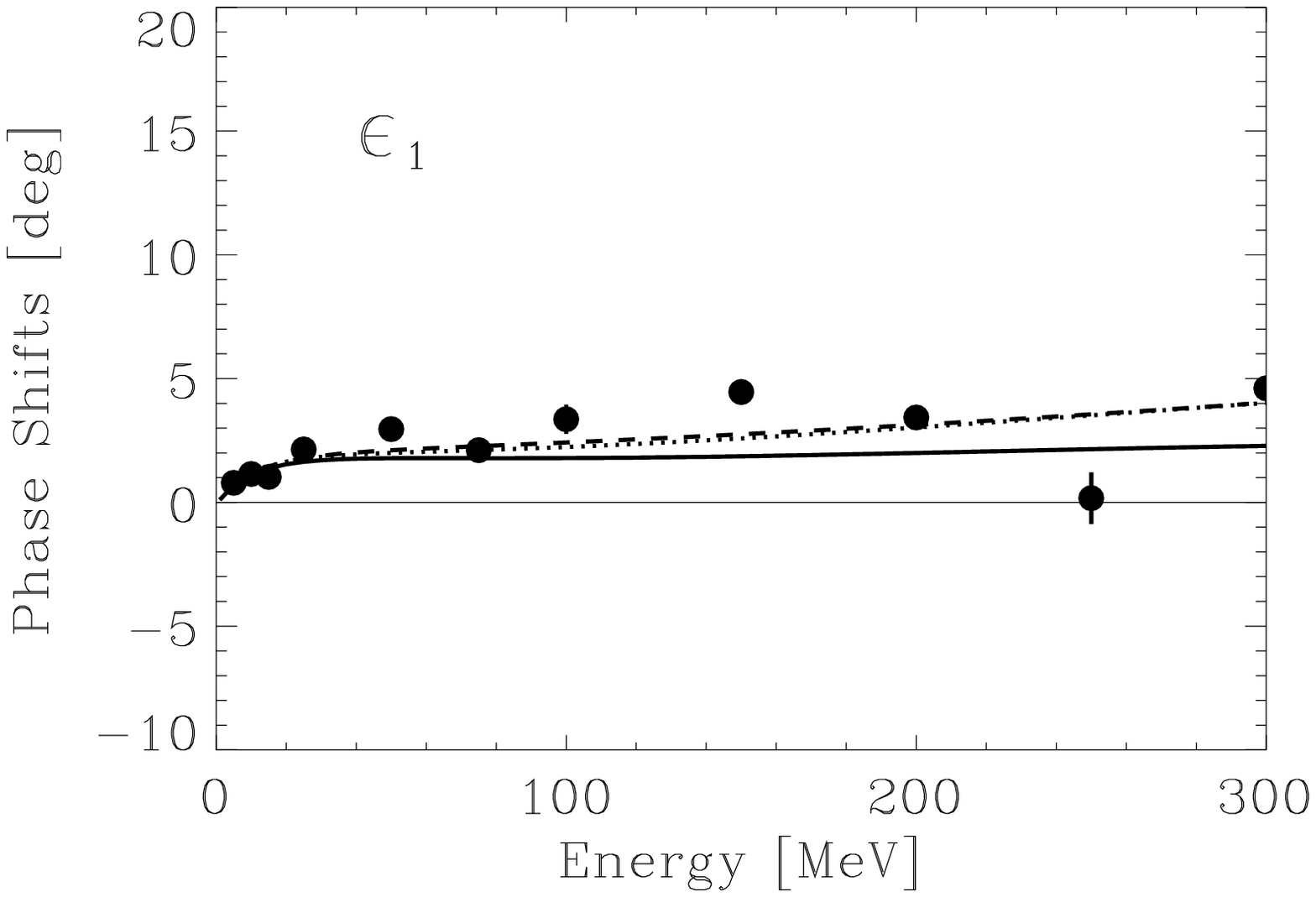,width=8.0cm}
\end{picture}
\begin{picture}(8,5.5)(0.0,0.0)
\epsfig{figure=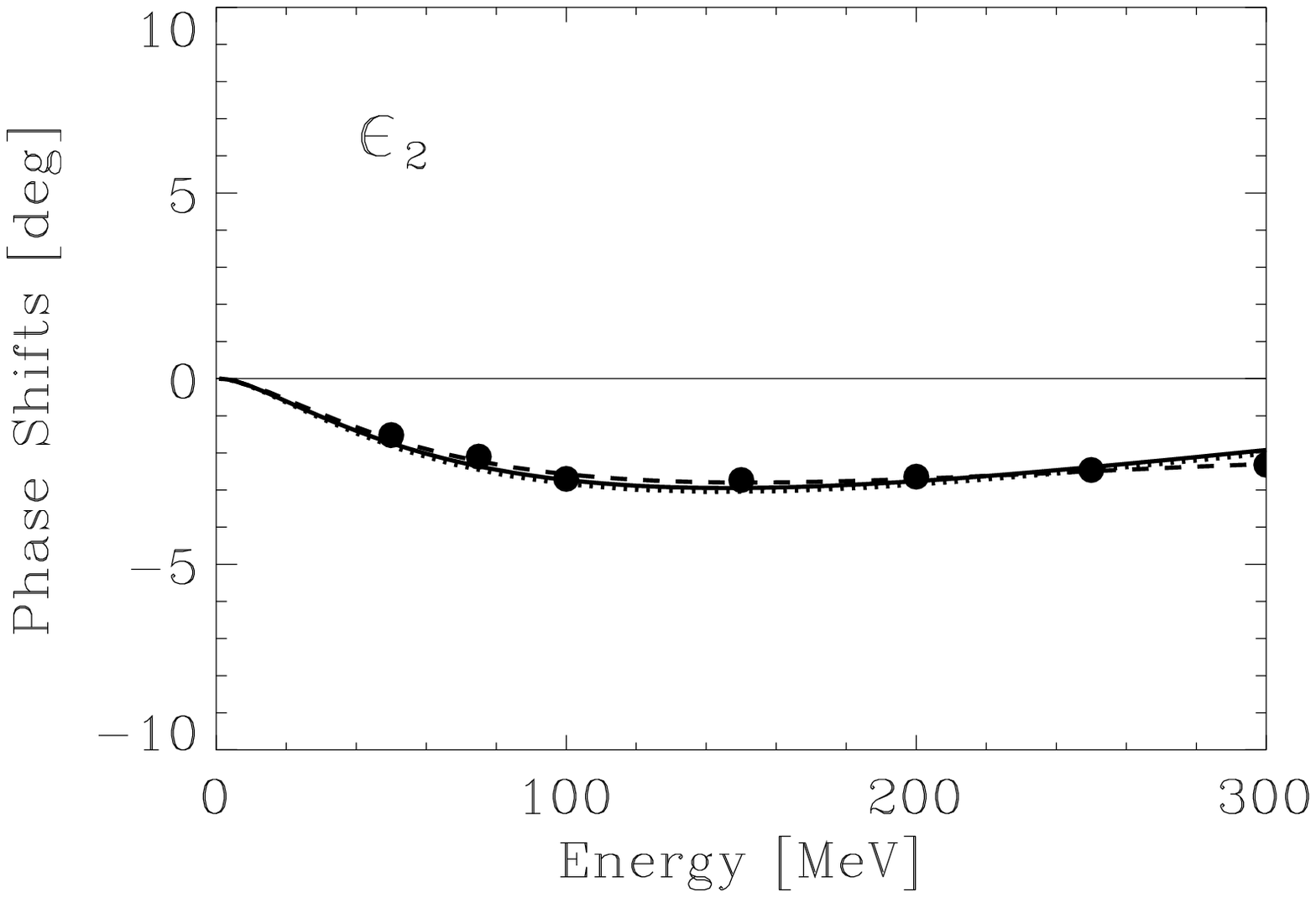,width=8.0cm}
\end{picture}
\caption[SYM $^3SD_1$ Phase Shifts]
{SYM Phase shifts for the coupled $^3SD_1$ and $^3PF_2$ channel.
\label{coupl}}
\end{figure}

To test our model we calculated $np$ phase shifts for total
angular momenta $J\le3$ and deuteron properties, neglecting 
meson-exchange-current contributions. 
Besides the meson masses in Table \ref{bbp} the values for the 
nucleon
mass and $\hbar c$ are the same as in the Bonn potential 
\cite{Mach87}
$$
M=938.926\;\mbox{MeV}\quad\mbox{and}\quad \hbar 
c=197.3286\;\mbox{MeV\,fm}.
$$
Most important we consistently applied the scaling law 
(\ref{scaling})
for the meson
self-interaction coupling constants. Thus 
the only adjustable parameters in the OSBEP are the pion 
self-interaction
coupling constant $\alpha_{\pi}$ and the meson-nucleon coupling
constants. For the $\pi${\it NN} coupling we used the 
experimentally fixed value 
$$
\frac{g_{\pi}^2}{4\pi}=13.7, 
$$
which was determined by the $\pi${\it N} and {\it NN} phase shift
analysis by Arndt {\it et al.}
\cite{gpi}. After all, our model includes eight parameters which
are adjusted to experiment. As experimental input the 
VPI-SM95 single energy phase shift
analysis \cite{SM95} and phases from the Nijmegen partial 
wave analysis (PWA) \cite{StSw} are available. 
Our model parameters were solely fitted to the
SM95 phase shifts which, in contrast to the Nijmegen PWA,
are given at single energies with individual error
bars. This allows one to calculate a measure of distance
($\chi^2$) for the model phase shifts with respect to the 
phases obtained from experiment. 
Table \ref{coco} contains the parameter set which produced the 
best agreement between model predictions and experimental data. 
Together with the Bonn B phase shifts \cite{Mach89} the phase
shifts from the OSBEP are shown in Figs.\,\ref{sc} and 
\ref{coupl}. 
Experimental values for the deuteron properties were taken from
different sources, they are shown in Table \ref{deut} 
compared with our theoretical results. 
\section{Conclusions and Outlook}
\label{cao}
To compare the OSBEP and Bonn B results we took the SM95
phase shifts and calculated (i) the distance to
the OSBEP, (ii) the distance to the Bonn B potential and (iii) 
the distance to the Nijmegen PWA. The respective numbers
($\chi^2$/datum) are 64.2 for the OSBEP, 92.2 for the Bonn B potential 
and 19.8 for the Nijmegen phase shifts. The slightly larger value
for the Bonn B potential arises since the parameters
therein were adjusted to the phase shifts available in the
1980s. Since the Nijmegen PWA
yields a $\chi^2$/datum with respect to the scattering
observables which is of order 1 (in their weighting of
data), the distance between Nijmegen PWA and SM95 is a
measure of consistency for the available phase shifts. If we
divide the distances of the OSBEP and the Bonn B potential 
by this number we
obtain a measure of consistency between the theoretical and
experimental data. The respective numbers are 3.2 for the
OSBEP and 4.7 for the Bonn B phase shifts. On this basis, the
agreement between OSBEP and experiment can be considered as
excellent, at least comparable to the Bonn B results. 
Additionally, the deuteron properties listed in Table
\ref{deut} coincide very well with the experimental data. 
These results were achieved by using eight
parameters only which enforces the evidence for the scaling law
(\ref{scaling}). 
   
We incorporated the nonlinear character of
chiral meson dynamics in a phenomenological way. Doing so the 
effect was
not only the possibility to substitute the form factors of 
conventional
models by a solitary meson propagator but additionally we found an
empirical scaling law to unify the meson self-interactions using a
single constant $\alpha_{\pi}$ only. The advantage of this 
result is a
more consistent description of the meson-nucleon coupling. 
Because of the
absence of form factors, the Ward identities for the conserved 
nucleon
currents are satisfied at each meson-nucleon vertex, a 
circumstance
which is not the
case in the Bonn B potential. The scaling law reflects a 
connection
between the mesons which may be traced to some fundamental 
symmetry on
the quark sector and therefore gives the parameter $\alpha_{\pi}$ a
strong physical significance. Knowing the 
astonishing influence of the meson self-interaction in {\it NN} 
interactions  
motivates a more detailed investigation concerning the 
connection between chiral symmetry and the nonlinear effects 
shown here. 
   
Clearly the model is a step towards a chiral symmetric 
boson-exchange
model but there is still a long way to go. First of all a realistic
chiral invariant Lagrangian must be constructed which 
contains all the
mesonic degrees of freedom of the boson-exchange model. 
The mechanism of spontaneous
symmetry breaking has to be worked out to explain the 
connection between
the meson parameters described by the scaling law. The 
possibility to
find such a Lagrangian, which yields excellent results in 
{\it NN} scattering
by construction when related to the nonlinear field equations 
used here, 
is an exciting challenge for future work. 
  
For nuclear reaction and nuclear structure calculations 
the solitary-boson-exchange potential can be applied in the 
same way as
the Bonn B potential. Furthermore, we expect that especially 
$\pi${\it N}
scattering or pion production
processes calculated in the framework of a model with solitary 
mesons
may show interesting results. Whereas the higher order terms 
($n>0$) in
the solitary pion propagator (\ref{psprop}) have only a small 
influence 
in {\it NN} interactions they can produce a major effect in pion 
production where
momenta are small and heavy mesons are kinematically suppressed.   
\acknowledgements
One of the authors (L.\,J.) would like to thank Prof.\ P.\,B. 
Burt from
the University of Clemson, Clemson, S.C., for hospitality and
instructive discussions.

\tighten

\begin{table} 
\caption{Bonn B meson parameters \protect\cite{Mach89}.}
\label{bbp}
\begin{tabular}{llllllll}
 & $\pi$ & $\eta$ & $\rho$ & $\omega$ & $\sigma_1$ &
$\sigma_0$ & $\delta$ \\
\hline
$S^P$ & $0^-$ & $0^-$ & $1^-$ & $1^-$ & $0^+$ & $0^+$ & $0^+$ \\

 $m_{\beta}$ (MeV) & 138.03 & 548.8 & 769 & 782.6 & 550 & 720 & 
983 \\
$\Lambda_{\beta}$ (GeV) & 1.7  & 1.5 & 1.85  & 1.85  & 1.9 & 2.0  
& 2.0 \\
 $n_{\beta}$  & 1 & 1  & 2  & 2  & 1 & 1 & 1 \\
\end{tabular}
\end{table}
\begin{table} 
\caption{Parameters from fitting the solitary meson
propagator to the Bonn B amplitude.}
\label{sicc}
\begin{tabular}{llllllll}
 & $\pi$ & $\eta$ & $\rho$ & $\omega$ & $\sigma_1$ & $\sigma_0$ 
&  $\delta$ \\\hline
$\;S^P\;$ & $0^-$ & $0^-$ & $1^-$ & $1^-$ & $0^+$ & $0^+$ 
& $0^+$ \\
 $\;m_{\beta}\;$ & 138.03 & 548.8 & 769 & 782.6 & 550 & 720 
& 983 \\
$\;\alpha_{\beta}\;$ & 0.36 & 0.1079 & 0.091 & 0.090 & 0.1755 
& 0.151 
& 0.1269 \\
\end{tabular}
\end{table}
\begin{table} 
\caption{OSBEP parameters.}
\label{coco}
\begin{tabular}{llllllll}
 & $\pi$ & $\eta$ & $\rho$ & $\omega$ & $\sigma_0$ & $\sigma_1$ 
& $\delta$ \\
\hline
$S^P$ & $0^-$ & $0^-$ & $1^-$ & $1^-$ & $0^+$ & $0^+$ & $0^+$ \\
 $\displaystyle{\frac{g_{\beta}^2}{4\pi}}$ &
    13.7   & 1.398  & 1.140 & 18.71 & 14.15 & 7.8389
& 1.369 \\
\multicolumn{4}{c}{$\alpha_{\pi}=0.428321$} &
\multicolumn{4}{c}{$f_{\rho}/g_{\rho}=4.422$} \\
\end{tabular}
\end{table}
\begin{table} 
\caption{Deuteron Properties}
\label{deut}
\begin{tabular}{lllll}
 & \multicolumn{1}{c}{Bonn B \protect\cite{Mach89}} &
\multicolumn{1}{c}{OSBEP} & 
\multicolumn{1}{c}{Exp.} &
\multicolumn{1}{c}{Ref.} \\
\hline
$E_B\mbox{ (MeV)}$ & $2.2246$ &    $2.224590$   
&  $2.22458900(22)$ &
\protect\cite{red}  \\   
$ \mu_d$           & $0.8514$ \tablenotemark[1]  &  
$0.8532$ \tablenotemark[1] &  $0.857406(1)$    & 
\protect\cite{rmu}  \\
$Q_d\mbox{ (fm$^2$)}$ &  $0.2783$ \tablenotemark[1] &
$0.2670$ \tablenotemark[1] &  $0.2859(3)$      &
\protect\cite{rqa}   \\
$A_S\mbox{ (fm$^{-1/2}$)}$  &  $0.8860$    &
   $0.8792$    &   $0.8802(20)$  &
\protect\cite{rqa}   \\
$D/S$           &   $0.0264$    &           
   $0.0256$    &   $0.0256(4)$   &
\protect\cite{rds}   \\
$r_{RMS}\mbox{ (fm)}$ &  $1.9688$     &      
  $1.9539$     &  $1.9627(38)$      &
\protect\cite{rqa}   \\
$P_D\quad(\%)$     &   $4.99$      &          
   $4.6528$      &
\multicolumn{1}{c}{--}    
&   \multicolumn{1}{c}{--}  \\
\end{tabular}
\tablenotetext[1]{Meson exchange current contributions not 
included}    
\end{table}
\clearpage
\widetext
%
%
%
\clearpage
%
%
%
%
\clearpage
%
%
\clearpage
%
%
\end{document}